\pdfoutput=1
\documentclass[10pt]{article}
\usepackage[preprint]{tmlr}

\usepackage{amsmath,amssymb}
\usepackage{booktabs}
\usepackage{graphicx}
\usepackage{xcolor}
\usepackage{hyperref}
\usepackage{url}
\usepackage{multicol}
\usepackage{chngcntr}   %
\usepackage{placeins}   %
\usepackage{makecell}
\usepackage{float}     %

\graphicspath{{figures/}}
\newcommand{\nModels}{15}
\newcommand{\nModelsRan}{16}
\newcommand{\nAnswers}{2.51M}
\newcommand{\parseMin}{98.5}
\newcommand{\parseMax}{100.0}
\newcommand{\exactMin}{12}
\newcommand{\exactMax}{34}
\newcommand{\catMin}{30}
\newcommand{\catMax}{63}
\newcommand{\slopeMin}{-0.21}
\newcommand{\slopeMax}{-0.48}
\newcommand{\crossMedian}{6.3}

\newcommand{\smallThresh}{15}
\newcommand{\nSmallModels}{11}

\newcommand{\partialModels}{\texttt{qwen3.6-27b}, \texttt{llama3.3-70b-awq} and \texttt{qwen2.5-72b-awq}}

\newcommand{\nPairs}{105}
\newcommand{\rMean}{0.827}

\newcommand{\pSameFam}{0.36}

\newcommand{\rSameGen}{0.871}
\newcommand{\rDiffGen}{0.824}
\newcommand{\pSameGen}{0.02}

\newcommand{\rSameFam}{0.833}
\newcommand{\rDiffFam}{0.825}

\newcommand{\backendName}{vllm}
\newcommand{\servTemp}{0.4}
\newcommand{\servTopP}{0.95}
\newcommand{\servMaxNew}{1024}

\newcommand{\nTeams}{4,928}

\newcommand{\bestHomoModel}{gpt-oss-20b}
\newcommand{\bestHomoMae}{1.628}

\newcommand{\predQuality}{+0.98}
\newcommand{\predCorr}{-0.10}

\newcommand{\rhoMedian}{0.922}

\newcommand{\idealPctMax}{10.7}

\newcommand{\idealPctMedian}{2.9}
\newcommand{\refRedMin}{51}
\newcommand{\refRedMax}{63}
\newcommand{\teamRedMax}{5.4}
\newcommand{\teamRedMed}{1.7}

\newcommand{\refItemRedMin}{74}
\newcommand{\refItemRedMax}{79}
\newcommand{\shareItemMed}{2.2}
\newcommand{\shareItemMax}{7.1}
\newcommand{\resampleGapMed}{+0.005}
\newcommand{\resampleGapMax}{+0.076}
\newcommand{\resampleGapMaxModel}{mistral-7b}
\newcommand{\nGainSig}{12}
\newcommand{\nGainModels}{15}
\newcommand{\rhoItemMedian}{0.906}
\newcommand{\shareRunMax}{11}
\newcommand{\nShareRun}{6}

\newcommand{\noAggGainModels}{\texttt{llama3.3-70b-awq}}

\newcommand{\bestMixTeam}{\texttt{gpt-oss-20b}+\texttt{qwen3.6-27b}}
\newcommand{\bestMixD}{+0.056}
\newcommand{\bestMixLo}{+0.010}
\newcommand{\bestMixHi}{+0.101}

\newcommand{\Bone}{0.211}
\newcommand{\Bthree}{0.014}
\newcommand{\Bcollapse}{93}

\newcommand{\zeroSpreadThree}{12}

\newcommand{\debateHelps}{8}
\newcommand{\debateHurts}{4}

\newcommand{\singleShot}{2.281}
\newcommand{\aggOnly}{2.251}
\newcommand{\withDebate}{2.233}
\newcommand{\totalGainPct}{2.1}

\newcommand{\misTeamPct}{22.5}

\newcommand{\totalGainPctDM}{2.2}

\newcommand{\scaffoldDelta}{-0.104}

\newcommand{\scaffoldPct}{4.6}

\newcommand{\scaffoldLo}{-0.196}
\newcommand{\scaffoldHi}{-0.015}
\newcommand{\logcalcDelta}{+0.540}

\newcommand{\logcalcCat}{66.9}
\newcommand{\gridCat}{51.5}

\newcommand{\heteroDebate}{+0.088}
\newcommand{\heteroDebateLo}{+0.011}
\newcommand{\heteroDebateHi}{+0.164}

\newcommand{\logcalcParse}{93.2}
\newcommand{\logcalcParseMin}{86.7}
\newcommand{\heteroDebateDM}{+0.012}
\newcommand{\heteroDebateLoDM}{-0.065}
\newcommand{\heteroDebateHiDM}{+0.092}
\newcommand{\peerMisPct}{32.6}

\newcommand{\BspreadTfour}{0.100}
\newcommand{\BspreadTten}{0.438}

\newcommand{\repRho}{0.021}
\newcommand{\repAgg}{0.065}

\newcommand{\auditFlagged}{29}
\newcommand{\auditItems}{529}
\newcommand{\auditRate}{5.5}
\newcommand{\auditOffsetMin}{1.5}
\newcommand{\auditSpreadMax}{1.0}

\newcommand{\nBiased}{10}
\newcommand{\nBiasedOf}{15}

\newcommand{\biasShrink}{1.00}
\newcommand{\biasShrinkIdeal}{3.87}

\newcommand{\nUnanimous}{4}
\newcommand{\nBands}{5}
\newcommand{\rIntraDraw}{0.906}
\newcommand{\rCrossDraw}{0.591}

\newcommand{\neffIntra}{1.10}
\newcommand{\neffCross}{1.65}
\newcommand{\rResidMin}{-0.010}
\newcommand{\rResidMax}{+0.032}
\newcommand{\rResidItemMax}{0.65}
\newcommand{\rResidItemMaxModel}{qwen3-8b}
\newcommand{\pooledMedAbs}{1.86}
\newcommand{\pooledMeanAbs}{3.65}

\newcommand{\maxAbsErr}{142}

\newcommand{\stableMaxRange}{0.11}

\newcommand{\rOneCapLo}{66}
\newcommand{\rOneCapHi}{85}

\newcommand{\nCompleteModels}{12}

\newcommand{\robMedAll}{1.860}
\newcommand{\robMedKeep}{1.783}
\newcommand{\robAall}{1.414}
\newcommand{\robAkeep}{1.339}

\newcommand{\robApct}{5.3}

\newcommand{\promptOldPct}{41}
\newcommand{\nPromptOldModels}{8}

\newcommand{\promptMixedModels}{\texttt{gpt-oss-20b}}
\newcommand{\promptLastOld}{21~July 2026}
\newcommand{\promptFirstNew}{23~July 2026}
\newcommand{\promptTokMin}{78}
\newcommand{\promptTokMax}{87}

\newcommand{\promptDmae}{-0.20}
\newcommand{\promptDmaeLo}{-0.24}
\newcommand{\promptDmaeHi}{-0.15}

\newcommand{\compTrue}{-0.021}
\newcommand{\compTrueLo}{-0.033}
\newcommand{\compTrueHi}{-0.010}
\newcommand{\compSelf}{-0.029}
\newcommand{\compSelfLo}{-0.043}
\newcommand{\compSelfHi}{-0.015}
\newcommand{\compPeerSelf}{+0.008}
\newcommand{\compPeerSelfLo}{-0.002}
\newcommand{\compPeerSelfHi}{+0.017}
\newcommand{\compBone}{0.221}
\newcommand{\compBtrueThree}{0.048}
\newcommand{\compBselfThree}{0.427}

\newcommand{\compPeerSelfDM}{+0.007}
\newcommand{\compPeerSelfDMLo}{-0.003}
\newcommand{\compPeerSelfDMHi}{+0.016}

\newcommand{\sciNBiased}{5}
\newcommand{\sciNBiasModels}{15}

\newcommand{\sciNNeg}{15}
\newcommand{\sciNFit}{15}

\newcommand{\sciRhoItem}{0.887}

\newcommand{\sciRefRedMin}{8}
\newcommand{\sciRefRedMax}{42}
\newcommand{\sciTeamRedMax}{5.7}
\newcommand{\sciTeamRedMed}{1.0}

\newcommand{\sciShareItemMed}{1.3}
\newcommand{\sciShareItemMax}{8.1}

\newcommand{\sciBcollapse}{84}
\newcommand{\sciGainPct}{5.5}
\newcommand{\sciPeerSelf}{+0.015}
\newcommand{\sciPeerSelfLo}{+0.003}
\newcommand{\sciPeerSelfHi}{+0.026}

\newcommand{\gueNModels}{14}

\newcommand{\gueRhoItem}{0.763}

\newcommand{\gueTeamRedMax}{21.7}
\newcommand{\gueTeamRedMed}{2.1}

\newcommand{\gueBcollapse}{61}
\newcommand{\gueGainPct}{8.0}
\newcommand{\guePeerSelf}{+0.003}
\newcommand{\guePeerSelfLo}{-0.033}
\newcommand{\guePeerSelfHi}{+0.054}

\newcommand{\gueIdealModel}{gpt-oss-20b}

\newcommand{\sciPeerSelfDM}{+0.007}
\newcommand{\sciPeerSelfDMLo}{-0.004}
\newcommand{\sciPeerSelfDMHi}{+0.018}

\newcommand{\gueNPos}{14}

\newcommand{\gueGtMin}{13}
\newcommand{\gueGtMax}{382}

\newcommand{\mcGainMainG}{2.1}

\newcommand{\mcTeamsSigMainM}{1}

\newcommand{\mcGainMainM}{1.7}

\newcommand{\mcTeamsSigRepM}{0}

\newcommand{\peerBudget}{120}
\newcommand{\homoGainFive}{0.035}
\newcommand{\divGainMed}{0.139}
\newcommand{\divGainPct}{6.3}

\newcommand{\divGainPos}{99.8}
\newcommand{\divBeatBest}{4.0}

\newcommand{\divMatched}{0.142}
\newcommand{\divMatchedLo}{0.095}
\newcommand{\divMatchedHi}{0.191}
\newcommand{\divMatchedPct}{6.6}
\newcommand{\divMatchedBest}{+0.035}
\newcommand{\divMatchedBestLo}{-0.046}
\newcommand{\divMatchedBestHi}{+0.117}

\newcommand{\rRobustMin}{0.64}
\newcommand{\rRobustMax}{0.67}
\newcommand{\rRobustResidMin}{0.60}
\newcommand{\rRobustResidMax}{0.69}

\newcommand{\rCoreItems}{486}
\newcommand{\trendShareMed}{17}
\newcommand{\trendShareMin}{6}
\newcommand{\trendShareMax}{29}
\newcommand{\trendShareCore}{11}
\newcommand{\recalSoloMed}{2.7}
\newcommand{\recalSoloMax}{11.8}
\newcommand{\recalTeamMed}{4.3}
\newcommand{\recalTeamMax}{12.9}
\newcommand{\recalTeamUncal}{1.5}
\newcommand{\recalTeamMaxModel}{\texttt{llama3.1-8b}}
\newcommand{\recalAMed}{+0.5}
\newcommand{\recalAMin}{-6.3}
\newcommand{\recalAMax}{+5.4}
\newcommand{\AvarOne}{2.64}
\newcommand{\AvarThree}{2.64}
\newcommand{\AvarChange}{-0.9}
\newcommand{\AvarChangeMin}{-5.1}
\newcommand{\AvarChangeMax}{+8.0}
\newcommand{\WvarChange}{-20}
\newcommand{\tPoolEight}{-0.013}
\newcommand{\tPoolEightLo}{-0.030}
\newcommand{\tPoolEightHi}{+0.004}
\newcommand{\tPoolEightMargin}{0.027}
\newcommand{\tPoolEightMarginPct}{1.4}

\newcommand{\tPoolTen}{-0.005}
\newcommand{\tPoolTenLo}{-0.024}
\newcommand{\tPoolTenHi}{+0.013}
\newcommand{\tPoolTenMargin}{0.021}
\newcommand{\tPoolTenMarginPct}{1.1}

\newcommand{\tPoolTenM}{-0.026}
\newcommand{\tPoolTenMLo}{-0.048}
\newcommand{\tPoolTenMHi}{-0.003}

\newcommand{\scaffoldP}{0.022}
\newcommand{\scaffoldHolm}{0.089}

\newcommand{\scaffoldDALo}{-0.89}
\newcommand{\scaffoldDAHi}{-0.10}
\newcommand{\scaffoldDApct}{-10}

\newcommand{\logcalcDB}{+6.45}
\newcommand{\steerChangeMin}{0.076}
\newcommand{\steerChangeMax}{0.112}

\newcommand{\floorChangeMin}{0.015}
\newcommand{\floorChangeMax}{0.052}

\newcommand{\skipDeltaMin}{-0.287}
\newcommand{\skipDeltaMax}{-0.241}
\newcommand{\skipRuns}{5}
\newcommand{\skipDApctMin}{-11}
\newcommand{\skipDApctMax}{-9}
\newcommand{\skipPct}{11}

\title{Shared Item-Level Error Limits the Aggregation of\\
       Small Language Models in Fermi Estimation}

\author{\name Carolina Fortuna \email carolina.fortuna@ijs.si\\
       \addr Jo\v{z}ef Stefan Institute, Ljubljana, Slovenia
       \AND
       \name Bla\v{z} Bertalani\v{c} \email blaz.bertalanic@ijs.si\\
       \addr Jo\v{z}ef Stefan Institute, Ljubljana, Slovenia}

\begin{document}
\maketitle

\begin{abstract}
Multi-agent systems of language models are often motivated by an averaging
argument: if the agents' errors are unbiased and independent, the error of an
aggregate of $N$ answers falls as $1/\sqrt{N}$. Fermi estimation, whose continuous
answers are aggregated on the logarithmic scale, permits a direct test. We decompose the $\log_{10}$ error of each answer into the mean error of
its model on the item, common to all its draws on the item, and a
residual, and estimate both from \nAnswers{} answers given before any
communication by \nModels{} open-weight models (3B--72B) on realFP,
with replication under a single prompt pair on realFP, Science Olympiad Fermi
questions and a guesstimation set. Given the item, teammates' errors are
uncorrelated and teams behave as independent draws, but the item-level
component accounts for a median $\rhoItemMedian$ of the error variance over
items. Teams of thirty therefore reduce a single agent's error by a median
\teamRedMed\% (at most \teamRedMax\%), whereas draws that were also unbiased on
each item would reduce it by \refItemRedMin--\refItemRedMax\%. All models
compress their estimates toward about $10^{6}$; this trend explains a median
\trendShareMed\% of the item-level component, and a cross-validated linear
recalibration on labelled items removes the overall bias but not that
component. Three rounds of debate reduce within-team dispersion by
\Bcollapse\% and leave the item-level component unchanged; revision without
peers attains the same small gain. Mixing five models of similar accuracy
reduces error by \divMatchedPct\% relative to their own homogeneous teams, but
no mixture improves on the most accurate model. Higher sampling temperature changes
team error by about 1\% at most, and contrastive activation steering not
detectably; per-agent scaffold prompts (not significant after multiplicity
correction) and skipping one layer in a single-model screen improve it. The results are conditional on the models,
benchmarks, prompts and settings tested.
\end{abstract}

\section{Introduction}
\label{sec:intro}

Systems that combine several language-model agents, by sampling and voting
\citep{wang2023selfconsistency,li2024moreagents}, by debate \citep{du2023debate}
or by committees of different models \citep{chen2024reconcile}, are often
motivated by an argument inherited from the wisdom of crowds
\citep{galton1907vox}: if the agents' errors are unbiased and independent, the
error of their aggregate falls as $1/\sqrt{N}$, and a larger team substitutes for
a better agent. Given its premises the argument is an identity, so its empirical
content lies in whether the premises hold. In forecast combination, positively
correlated errors are known to limit the benefit of averaging
\citep{clemen1986combining}, and recent work reports correlated errors across
language models on discrete benchmarks \citep{kim2025correlated,goel2025great}.

Fermi estimation permits a quantitative test. Fermi problems ask for quantities
that can only be estimated \citep{kalyan2021fermi}; their answers are
continuous, are naturally aggregated by a mean on the logarithmic scale, and
their errors can cancel rather than merely outvote one another. The error of an
answer can therefore be measured in orders of magnitude (OOM) and decomposed. We
write the $\log_{10}$ error of each answer as the mean error of its model on the
item, which every draw of that model on that item shares, plus a residual, and
estimate both components from \nAnswers{} answers that \nModels{} open-weight
models (3B--72B parameters, \nSmallModels{} of them below \smallThresh{}B) gave
to the 529 items of realFP before any communication, in teams of 2 to 30
agents. A second corpus with a single prompt pair repeats the analyses on realFP,
on Science Olympiad Fermi questions (SciOly) and on a guesstimation set
\citep{chuang2025woc}.

The estimands are defined over the items of a benchmark
(Section~\ref{sec:framework}). The item-level component measures on which items,
and by how much, all draws of a model err; it is not a dependence between agents
on a fixed item, which we find to be absent. We ask four questions; each answer
is conditional on the models, benchmarks, prompts and decoding settings tested.
\begin{itemize}
\item \textbf{RQ1: How is single-agent error structured?}
(Section~\ref{sec:round1}) Given the item and run, the residuals of teammates
are uncorrelated, but the item-level component accounts for a median
$\rho=\rhoItemMedian$ of the error variance over items. All \nModels{} models
compress magnitudes; this trend explains a median \trendShareMed\% of the
item-level component, and the per-item errors of different models remain
correlated at $\rRobustResidMin$--$\rRobustResidMax$ after it is removed.
\item \textbf{RQ2: How much does aggregation recover?} (Section~\ref{sec:agg})
Real teams have the error of teams resampled independently from each item's
draws, to within a median $\resampleGapMed$~OOM. They reduce single-agent error
by a median \teamRedMed\%, which is \shareItemMed\% of the reduction that draws
unbiased and independent given the item would deliver.
\item \textbf{RQ3: Does communication reduce the item-level error?}
(Section~\ref{sec:debate}) No. Debate reduces within-team dispersion by
\Bcollapse\% and changes the item-level component by a median $\AvarChange$\%;
revision without peers obtains at least the same small gain.
\item \textbf{RQ4: Which interventions reduce it?}
(Section~\ref{sec:interventions}) Mixing models of similar accuracy reduces
error by \divMatchedPct\% relative to their own teams, without improving on the
best of them; temperature changes team error by about 1\% at most and
activation steering not detectably; per-agent scaffold prompts and skipping one
layer improve the team, the first without surviving correction for multiplicity
and the second in a 150-item screen of one model. A linear recalibration on labelled items removes the overall bias and
leaves the item-level component unchanged.
\end{itemize}

\section{Related work}
\label{sec:related}

\paragraph{Combining judgements.}
The benefit of averaging depends on the dependence of the errors averaged:
positively correlated forecasts carry less information than their number
suggests \citep{clemen1986combining}, and the effective number of independent
estimators under an intraclass correlation $\rho$ is the design effect of
\citet{kish1965survey}. Averaging repeated estimates of one person, the ``crowd
within'', helps less than averaging different people \citep{vul2008crowd}, and
asking a person to reconsider from a different premise increases the gain
\citep{herzog2009wisdom}; homogeneous teams and per-agent prompts are their
analogues for language models. Social influence makes estimates converge without
making them more accurate \citep{lorenz2011social}, with effects that depend on
the network of influence \citep{becker2017network}.

\paragraph{Aggregating language models.}
Sampling several reasoning paths and voting improves accuracy on discrete
reasoning tasks \citep{wang2023selfconsistency}, with gains that grow with the
number of agents \citep{li2024moreagents}. Multi-agent debate was proposed to
improve factuality and reasoning \citep{du2023debate}, but its gains depend on
the protocol \citep{smit2024mad} and are largely attributable to the vote over
agents rather than to the exchange \citep{choi2025debate}; without external
feedback, self-correction does not reliably improve reasoning
\citep{huang2024selfcorrect}. Errors of different language models are
correlated, more so for more capable models
\citep{kim2025correlated,goel2025great}, and several studies identify the
correlation between agents as the limit of multi-agent systems
\citep{chen2024reconcile,yang2026diversity,patel2026collapse}. These studies
measure agreement on discrete outputs or in representation space; on a
continuous task the common component can be measured in the units of the answer.

\paragraph{Fermi estimation and numerical magnitude.}
Single models score low on realFP \citep{kalyan2021fermi,liu2025fermistudy} and
are overconfident in their intervals \citep{epstein2025fermieval}.
\citet{chuang2025woc} aggregate 30 reasoning paths of one model by the median and
report large gains for weak models and none for the strongest. Human magnitude
judgements regress toward the centre of the stimulus range
\citep{hollingworth1910central,petzschner2015bayesian}, and language models
encode numbers on a compressed scale \citep{alquboj2025numbers,cacioli2026weber};
we measure a behavioural counterpart of that compression and its consequence for
aggregation.

\section{Framework and data}
\label{sec:framework}

\subsection{Decomposition and estimands}
\label{sec:decomp}

Fix a model and a benchmark. Let $e_{jk}=\log_{10}\hat y_{jk}-\log_{10}y_j$ be
the signed error, in OOM, of draw $k$ on item $j$ with true value $y_j$. We write
\begin{equation}
e_{jk} = \mu + c_j + u_{jk}, \qquad \textstyle\mathbb{E}_j\, c_j = 0,\qquad
\mathbb{E}[u_{jk}\mid j]=0,
\label{eq:decomp}
\end{equation}
where $\mu$ is the model's overall bias, $\mu+c_j$ its mean error on item $j$,
which all of its draws on that item share, and $u_{jk}$ the residual of one
draw. Moments are taken over the items of the benchmark and the draws on each
item:
\begin{equation}
A=\operatorname{Var}_j(c_j),\qquad B=\mathbb{E}_j\operatorname{Var}(u_{jk}\mid j),\qquad
\rho=\frac{A}{A+B},\qquad N_{\mathrm{eff}}(N)=\frac{N}{1+(N-1)\rho}.
\label{eq:rho}
\end{equation}
$A$ is the between-item variance of the item-mean error, $B$ the average
within-item variance, $\rho$ the intraclass correlation of draws over items, and
$N_{\mathrm{eff}}(N)$ the number of independent draws that have the same mean
squared error over items as a team of $N$ \citep{kish1965survey}.

\paragraph{What aggregation can remove.}
A team combines its members' draws on item $j$ by a rule $g$. For every rule
that shifts with its inputs, $g(x+a)=g(x)+a$, such as the mean, the median or a
trimmed mean, the team error is $\mu+c_j+g(u_{j1},\dots,u_{jN})$: the
item-level error passes through unchanged and only the residuals are combined.
If the draws are independent given the item, the mean of $N$ of them has
expected squared error $\mu^2+A+B/N$ over items, and $\mu^2+A$ is a floor for
every such rule. It is not a floor for procedures that use information beyond
the answers, such as a recalibration learned on labelled items, or for
interventions that change the agents themselves; both can change $\mu$ and $c_j$
(Section~\ref{sec:interventions}). $\rho$ and $A$ describe a benchmark's item
distribution, not dependence between agents given the item, which we test
separately (Section~\ref{sec:dependence}).

\paragraph{Estimation.}
$A$ and $B$ are estimated by a one-way decomposition by item of the round-1
draws of a model, with errors winsorized at $\pm5$ OOM and $A$ corrected for the
sampling variance of the item means; a fit of the variance of real team means,
$\operatorname{Var}(\bar e_N)=A+B/N$, gives a second estimate
(Appendix~\ref{app:agg}). For debate we also report the within-team variance
$W$, and the median over items of the within-team interquartile range, which we
call the \emph{dispersion} $D$. A degraded agent inflates $B$ and thereby lowers
$\rho$ without any change in $A$, so $\rho$ is never used alone as evidence of
diversity; an intervention is credited with reducing the item-level error only
if it lowers $A$ \emph{and} improves the team.

\subsection{References for aggregation}
\label{sec:refs}

The rate $1/\sqrt{N}$ describes the standard error of a mean; for the
winsorized absolute error of a geometric mean of heavy-tailed errors it does not
hold. We therefore simulate two references per model and team size, scored by
the same rule and metric as the real teams. \emph{Resampled given the item}: each
member is drawn independently, with replacement, from the model's round-1 errors
on the same item (about 350 per item). This is the counterfactual of members that
are independent given the item; if the real teams match it, their shortfall is
not due to dependence between members. \emph{Unbiased given the item}: the same
draws with independent random signs, which removes $\mu+c_j$ while keeping each
item's error magnitudes. This is the reduction that aggregation would deliver if
the draws were also unbiased on each item, and it is our reference. A third
reference, which draws members across items and therefore also removes item
difficulty, is reported in Appendix~\ref{app:agg}.

\subsection{Data and protocol}
\label{sec:protocol}

\begin{table}[t]
\centering\footnotesize
\setlength{\tabcolsep}{3.5pt}
\begin{tabular}{lllllll}
\toprule
corpus & models & items & $N$ & rounds & prompts & used in \\
\midrule
main sweep & \nModels{} & realFP 529 & 2--30 & 1--3, genuine peers & both pairs & \S\ref{sec:round1}--\ref{sec:debate} \\
second corpus & \nModels{} & realFP, SciOly 500, Guess.\ 15 & 2--30 & 1--3, peers and self-revision & revised & \S\ref{sec:round1}--\ref{sec:debate} \\
heterogeneous pool & 5 & realFP 529 & 3, 5 & 1--3, three peer sources & revised + arms & \S\ref{sec:debate}, \ref{sec:interventions} \\
temperature sweep & 4 & realFP 529 & 10 & 1 & revised & \S\ref{sec:interventions} \\
activation screen & 1 & realFP 150 & 5 & 1 & revised & \S\ref{sec:interventions} \\
\bottomrule
\end{tabular}
\caption{Corpora. All sample at $T=\servTemp$ (the temperature sweep also at 0.8
and 1.0) with top-$p=\servTopP$, \servMaxNew{} new tokens (512 in the activation
screen) and thinking modes disabled. ``Both pairs'': the main sweep mixes two
prompt pairs (Section~\ref{sec:protocol}). Guesstimate has answers from 14
models. Models, serving and coverage: Appendix~\ref{app:models}.}
\label{tab:corpora}
\end{table}

\paragraph{Benchmarks.}
Of the 557 realFP test questions \citep{kalyan2021fermi}, 529 have a positive
finite numeric answer, from $10^{-29}$ to $10^{90}$, with the middle half within
$10^{1.8}$--$10^{7.6}$. The 500 SciOly test
questions\footnote{\url{https://github.com/landy8697/open-scioly-fermi}, split
as in \citet{epstein2025fermieval}.} record only the power of ten of each answer,
$y=10^k$; an exactly correct estimate $x\cdot 10^k$ then has
$e=\log_{10}x\in[0,1)$, an upward offset common to all agents on an item, which
enters $\mu$ and $c_j$ but not $B$. The 15 Guesstimate questions of
\citet{chuang2025woc} count objects in containers (13 to 382); with so few items,
spanning less than two orders of magnitude, we draw no conclusion from them
alone.

\paragraph{Models and protocol.}
\nModelsRan{} instruction-tuned open-weight models from nine lineages ran on the
task (Table~\ref{tab:models}); \texttt{r1-distill-qwen-14b} reaches the token
limit on most generations and is excluded, leaving \nModels{}, of which the two
largest are AWQ 4-bit checkpoints. In round~1 every member of a team receives the
same prompt, without exemplars or peer content, so a homogeneous team is $N$
samples from one predictive distribution; in rounds 2 and 3 agents see their
peers' answers and revise. The prompts (Appendix~\ref{app:prompts}) request
explicit reasoning. The main sweep mixes an original and a revised prompt pair
(\promptOldPct\% of its round-1 answers, unevenly across \nPromptOldModels{}
models), which we identify from prompt length (Appendix~\ref{app:models}); the
revised pair lowers MAE@5 by $\promptDmae$~OOM $[\promptDmaeLo, \promptDmaeHi]$
in the models that ran both. The second corpus, which uses the revised pair
throughout, reproduces every structural statistic of
Sections~\ref{sec:round1}--\ref{sec:agg} (Table~\ref{tab:bench}); levels of
accuracy and bias of individual models carry this qualification.

\subsection{Error metric and inference}
\label{sec:metric}

An answer is correct to the order of magnitude if $|e|\le0.5$ and catastrophic
if $|e|>2$. Errors are heavy-tailed (pooled median $|e|$ \pooledMedAbs, mean
\pooledMeanAbs, maximum \maxAbsErr~OOM), so the headline metric is the mean
absolute error winsorized at 5 OOM, $\mathrm{MAE@5}=\frac1n\sum\min(|e|,5)$.
The team estimate is the geometric mean of the members' estimates, so its signed
error is the mean of theirs and Equation~\ref{eq:decomp} applies directly;
Appendix~\ref{app:teamrule} repeats every team-level result with the median of
the $\log_{10}$ estimates. Each model answers each item about 350 times, so the
item is the unit of inference, and intervals are 95\% item-clustered bootstrap
intervals. One family of contrasts is confirmatory: the six prompt-level arms of
Section~\ref{sec:interventions}, whose endpoint (the paired difference in team
MAE@5) was fixed in an internal analysis plan before the arms were scored; the
plan was not registered publicly. We report Holm-adjusted $p$-values for this
family. All other tests are exploratory, and counts of models with a
significant effect are descriptive and unadjusted.

\section{Structure of single-agent error (RQ1)}
\label{sec:round1}

\begin{figure}[t]
\centering
\includegraphics[width=0.9\textwidth]{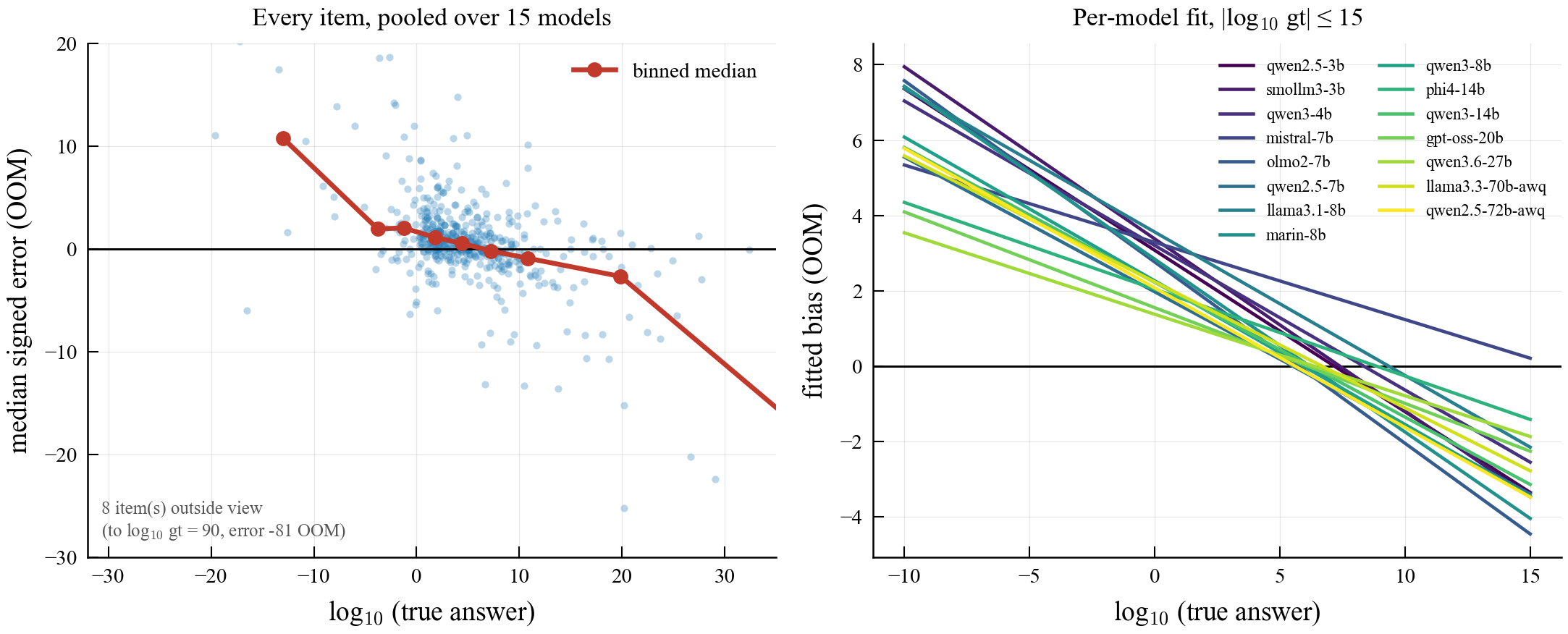}
\caption{Round-1 estimates compress toward a central magnitude. Left: per-item
median signed error against the true answer, pooled over \nModels{} models, with
the binned median. Right: per-model linear fits on $|\log_{10}y|\le15$.}
\label{fig:compression}
\end{figure}

\paragraph{Accuracy and magnitude compression.}
A single agent is correct to the order of magnitude on \exactMin--\exactMax\% of
realFP items and catastrophically wrong on \catMin--\catMax\%, with parse rates
of \parseMin--\parseMax\% (Table~\ref{tab:r1summary}). Every model overestimates
small quantities and underestimates large ones (Figure~\ref{fig:compression}).
Regressing the per-item median signed error on $\log_{10}y$, excluding the 43
items with $|\log_{10}y|>15$, gives a negative slope for all \nModels{} models,
from $\slopeMin$ to $\slopeMax$, with a median crossing point at
$10^{\crossMedian}$; the four checkpoints above 15B lie inside this range. On
SciOly \sciNNeg{} of \sciNFit{} slopes are negative, and the Guesstimate
quantities, which lie below the crossing point, are overestimated by \gueNPos{}
of \gueNModels{} models. The compression is common to all models and is
therefore part of $c_j$, but a minor part: the linear trend explains a median
\trendShareMed\% (range \trendShareMin--\trendShareMax\%) of $A$, and
\trendShareCore\% on $|\log_{10}y|\le15$.

\paragraph{Bias.}
\nBiased{} of \nBiasedOf{} models have a mean signed error whose interval
excludes zero, and every point estimate is positive in the main sweep (14 of 15
under the revised pair alone). Because the biases share a sign, the bias of a
consensus of $M$ models equals the mean bias of its members and does not shrink
with $M$ (shrinkage \biasShrink$\times$, against $\biasShrinkIdeal\times$ for
independent biases), and in \nUnanimous{} of \nBands{} magnitude bands every
model errs in the same direction (Appendix~\ref{app:round1}). On SciOly the
answer key offsets all signed errors upward by an unknown common amount, so its
count of biased models (\sciNBiased{} of \sciNBiasModels{}) is a lower bound and
is not comparable with realFP.

\subsection{Dependence given the item and correlation over items}
\label{sec:dependence}

The independence premise can fail in two senses. \emph{Given the item}, two
members of a team could share more than the item's mean error; this is the
dependence that averaging requires to be absent. The residuals of two teammates,
measured against the mean of their item and run, correlate at between
$\rResidMin$ and $\rResidMax$ across models, as independent sampling implies.
Measured against the item mean alone they correlate at up to $\rResidItemMax$
(\texttt{\rResidItemMaxModel}), because some models' runs used different prompt
versions and the members of a team share their run's deviation; this run
component accounts for up to \shareRunMax\% of the error variance in
\nShareRun{} models.

\emph{Over items}, two draws correlate because both carry $c_j$. Two draws of
the same model correlate at $\rIntraDraw$, which equals the one-way estimate of
$\rho$ (\rhoItemMedian) and is close to the team-level fit ($\rhoMedian$, which
also counts the run component); $N_{\mathrm{eff}}(30)$ is then \neffIntra.
Draws of different models correlate at $\rCrossDraw$
($N_{\mathrm{eff}}(30)=\neffCross$). The per-item mean errors of two models
correlate at $\rMean$ on average over the \nPairs{} pairs; this Pearson
correlation is inflated by the items of extreme magnitude, and robust variants
(Spearman, errors winsorized at 5~OOM, or the \rCoreItems{} items with
$|\log_{10}y|\le15$) give $\rRobustMin$--$\rRobustMax$, and
$\rRobustResidMin$--$\rRobustResidMax$ with the linear magnitude trend removed.
Models of the same family are not distinguishably more correlated than others
(Appendix~\ref{app:round1}).

These are correlations of errors: the models err on the same items in the same
direction beyond the common magnitude trend. They do not identify the cause.
Shared knowledge acquired from similar text, properties of the items and their
construction, and errors in the answer key would all produce them. An audit of
the answer key flags \auditRate\% of items as candidate label errors, and
removing them changes the cross-model mean error per item by \robApct\%
(Appendix~\ref{app:round1}), so label errors are not the main source.

\begin{table}[t]
\centering\footnotesize
\setlength{\tabcolsep}{4pt}
\resizebox{\textwidth}{!}{\begin{tabular}{p{4.7cm}cccc}
\toprule
 & \makecell{realFP\\(main)} & \makecell{realFP\\(second)} & \makecell{SciOly\\(second)} & \makecell{Guesstimate\\(second)} \\
\midrule
models & 15 & 15 & 15 & 14 \\
items & 529 & 529 & 500 & 15 \\
one agent within an order of magnitude (\%) & 12--34 & 14--34 & 15--46 & 3--67 \\
one agent off by more than $100\times$ (\%) & 30--63 & 31--57 & 17--54 & 0--80 \\
models with mean error $\neq 0$ & 10/15 & 9/15 & 5/15 & 12/14 \\
bias shrinkage across models (realised / $\sqrt{M}$) & 1.00 / 3.87 & 1.04 / 3.87 & 1.65 / 2.83 & 1.00 / 3.74 \\
models with a negative compression slope & 15/15 & 15/15 & 15/15 & 12/14 \\
crossing point (median, $\log_{10}$) & 6.3 & 6.4 & 6.6 & 2.9 \\
cross-model $r$ of per-item errors (raw / magnitude removed) & 0.827 / 0.689 & 0.827 / 0.690 & 0.689 / 0.581 & 0.407 / 0.395 \\
single-draw $r$ (same model / different models) & 0.906 / 0.591 & 0.913 / 0.601 & 0.887 / 0.557 & 0.733 / 0.368 \\
$\rho$, one-way by item (median over models) & 0.906 & 0.915 & 0.887 & 0.763 \\
$\rho$, fit to team means (median over models) & 0.922 & 0.920 & 0.890 & 0.758 \\
$N_{\mathrm{eff}}(30)$ (median over models) & 1.08 & 1.08 & 1.12 & 1.31 \\
team of 30 $-$ independent draws given the item (OOM; median) & $+0.005$ & $+0.000$ & $+0.000$ & $+0.004$ \\
error reduction at $N{=}30$, real teams (\% of one agent; median / max) & 1.7 / 5.4 & 1.7 / 6.0 & 1.0 / 5.7 & 2.1 / 21.7 \\
\quad models whose reduction excludes zero & 12/15 & 11/15 & 6/15 & 9/14 \\
error reduction at $N{=}30$, draws unbiased and independent given the item (\%; range) & 74--79 & 74--79 & 57--76 & 81--85 \\
share of that reduction realised (\%; median / max) & 2.2 / 7.1 & 2.2 / 7.8 & 1.3 / 8.1 & 2.5 / 26.8 \\
mixtures significantly better than the best homogeneous team & 0/4,928 & 0/4,928 & 0/4,928 & 0/3,458 \\
debate: reduction of within-team dispersion $D$ at $N{=}5$ (\%) & 93 & 78 & 84 & 61 \\
debate: total gain over one agent (\%) & 2.1 & 2.4 & 5.5 & 8.0 \\
round 3: genuine peers $-$ self-revision (OOM) & -- & $+0.008$ $[-0.002, +0.017]$ & $+0.015$ $[+0.003, +0.026]$ & $+0.003$ $[-0.033, +0.054]$ \\
\bottomrule
\end{tabular}}
\caption{Principal statistics on the four data sets, computed by the same code
with the same error, team rule (geometric mean) and thresholds. ``Main'': the
main sweep; ``second'': the second corpus, which uses the revised prompt pair
throughout. Counts of models are descriptive and unadjusted. SciOly errors carry
the offset of its answer key; with 15 items, the slopes, correlations and $\rho$
of Guesstimate are imprecise. Per-model results:
Appendices~\ref{app:round1}--\ref{app:bench}.}
\label{tab:bench}
\end{table}

\section{Aggregation (RQ2)}
\label{sec:agg}

Round~1 of every team is an aggregation experiment: $N$ answers are produced
before any communication, for $N\in\{2,3,5,7,10,15,20,25,30\}$, in three runs on
all 529 items. We aggregate these real teams by the geometric mean, without
resampling, and compare each team with the individual agents of the same cell.

\begin{figure}[t]
\centering
\includegraphics[width=0.9\textwidth]{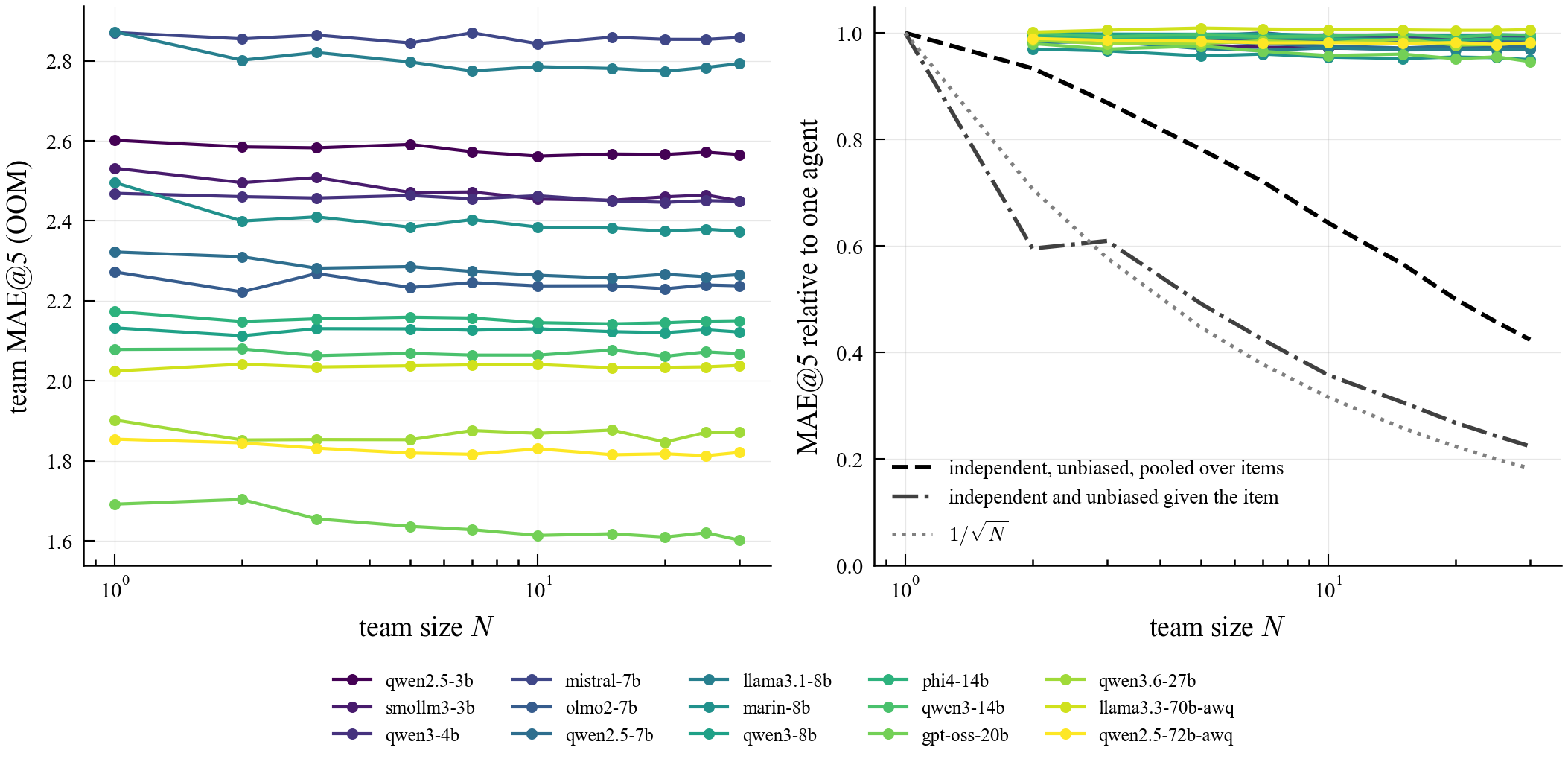}
\caption{Left: team MAE@5 against $N$ (realFP, main sweep). Right: the same
relative to one agent, with the median over models of three references: draws
independent and unbiased given the item (dash-dot; the kink at $N=2$ arises
because identical draws with opposite signs cancel exactly), draws independent
and unbiased across items (dashed), and the $1/\sqrt{N}$ rate (dotted).}
\label{fig:agg}
\end{figure}

\paragraph{Teams behave as independent draws given the item.}
The real teams have nearly the error of teams resampled independently from each
item's draws: at $N=30$ the median difference is $\resampleGapMed$~OOM, and the
largest, $\resampleGapMax$ for \texttt{\resampleGapMaxModel}, occurs in a model
whose runs mixed prompt versions; in the second corpus the median difference is
below 0.001~OOM (Table~\ref{tab:bench}). The shortfall of aggregation is
therefore not due to dependence between members but to the error that each
item imposes on all of them.

\paragraph{The reduction realised.}
Teams of thirty reduce single-agent error by a median \teamRedMed\% and at most
\teamRedMax\% (Figure~\ref{fig:agg}); the reduction is distinguishable from zero
for \nGainSig{} of \nGainModels{} models, and for \noAggGainModels{} the team is
not better than one agent. Draws unbiased and independent given the item would
reduce the error by \refItemRedMin--\refItemRedMax\%, so the real teams realise a
median \shareItemMed\% (at most \shareItemMax\%) of that reduction. The
conclusion does not depend on the construction of the reference: against
members drawn across items the available reduction is \refRedMin--\refRedMax\%
and the share realised at most \idealPctMax\% (Appendix~\ref{app:agg}). In mean
squared error over items, a team of thirty is worth $N_{\mathrm{eff}}(30) =
\neffIntra$ independent agents.

\paragraph{Other benchmarks.}
The second corpus reproduces this on realFP and SciOly (Table~\ref{tab:bench}):
on SciOly teams of thirty reduce single-agent error by a median
\sciTeamRedMed\% (at most \sciTeamRedMax\%), \sciShareItemMed\% of the reduction
available under the per-item reference. On Guesstimate one model,
\texttt{\gueIdealModel}, reduces its error by \gueTeamRedMax\%, the only
reduction above one tenth in the study; the median model reduces it by
\gueTeamRedMed\%. This benchmark has the lowest $\rho$ (\gueRhoItem), that is, the
largest within-item share of the error, which is the condition under which the
decomposition predicts a gain; it is also where \citet{chuang2025woc} report
large gains from aggregating sampled estimates.

\section{Communication between agents (RQ3)}
\label{sec:debate}

In rounds 2 and 3 each agent sees its teammates' answers and revises. Round~1
of the same teams serves as a within-subject control.

\paragraph{Dispersion collapses; the item-level error does not change.}
At $N=5$, from round~1 to round~3, the dispersion $D$ falls from \Bone{} to
\Bthree{}~OOM ($-\Bcollapse$\%), and by round~3, \zeroSpreadThree{} of
\nModels{} models have $D=0$: on at least half of the items, three of five
agents return the same number. The within-team variance $W$, which occasional
outlying members dominate, falls by a median $\WvarChange$\%. The item-level
component is unchanged: $\sqrt{A}$ is \AvarOne{}~OOM at round~1 and
\AvarThree{}~OOM at round~3, a median change in $A$ of $\AvarChange$\% (range
$\AvarChangeMin$ to $\AvarChangeMax$\%; Table~\ref{tab:rounds}). Team error
falls from \singleShot{}~OOM for one agent to \aggOnly{} for five aggregated
agents and \withDebate{} after debate, a total of \totalGainPct\% at fifteen
times the generation cost; per model, debate improves the team for
\debateHelps{} and worsens it for \debateHurts{} of \nModels{}. On SciOly and
Guesstimate debate reduces $D$ by \sciBcollapse\% and \gueBcollapse\% and total
error by \sciGainPct\% and \gueGainPct\%.

\begin{table}[t]
\centering\small
\begin{tabular}{lrrrr}
\toprule
 & \multicolumn{2}{c}{heterogeneous pool (529 items)} & \multicolumn{2}{c}{homogeneous teams (529 items)} \\
\cmidrule(lr){2-3}\cmidrule(lr){4-5}
contrast & $\Delta$ & 95\% CI & $\Delta$ & 95\% CI \\
\midrule
genuine peers, round 3 $-$ round 1 & $\mathbf{+0.088}$ & $[+0.011, +0.164]$ & $\mathbf{-0.021}$ & $[-0.033, -0.010]$ \\
self-revision, round 3 $-$ round 1 & $+0.057$ & $[-0.003, +0.116]$ & $\mathbf{-0.029}$ & $[-0.043, -0.015]$ \\
peers from other items, round 3 $-$ round 1 & $\mathbf{+0.187}$ & $[+0.111, +0.261]$ & \multicolumn{2}{c}{--} \\
round 3: genuine peers $-$ other items & $\mathbf{-0.097}$ & $[-0.187, -0.006]$ & \multicolumn{2}{c}{--} \\
round 3: genuine peers $-$ self-revision & $+0.033$ & $[-0.045, +0.113]$ & $+0.008$ & $[-0.002, +0.017]$ \\
\bottomrule
\end{tabular}
\caption{Change in team MAE@5 (OOM), paired by item, with 95\% item-clustered
intervals; positive values indicate degradation, bold an interval that excludes
zero. Left: baseline arm of the heterogeneous pool. Right: second corpus, pooled
over \nModels{} models with equal weight and over all team sizes; it ran no
placebo arm.}
\label{tab:peers}
\end{table}

\paragraph{Revision without peers.}
Round~3 differs from round~1 by exposure to peers and by two further generation
passes. The second corpus ran, for every model, a self-revision arm in which the
agent revises seeing only its own previous answer (Table~\ref{tab:peers}). On
realFP team error changes from round~1 to round~3 by $\compSelf$
$[\compSelfLo, \compSelfHi]$~OOM under self-revision and by $\compTrue$
$[\compTrueLo, \compTrueHi]$ with peers; at round~3 the difference,
$\compPeerSelf$ $[\compPeerSelfLo, \compPeerSelfHi]$, favours self-revision
without excluding zero (under the median rule it does exclude zero,
Appendix~\ref{app:teamrule}). On SciOly the difference is $\sciPeerSelf$
$[\sciPeerSelfLo, \sciPeerSelfHi]$ and on Guesstimate $\guePeerSelf$
$[\guePeerSelfLo, \guePeerSelfHi]$. The protocols differ most in their effect on
disagreement: with peers $D$ falls from \compBone{} to \compBtrueThree{}~OOM,
whereas under self-revision it rises to \compBselfThree{}. In these data,
convergence is produced by exposure to peers and the small gain by the revision
passes, consistent with reports that the exchange between agents adds little
beyond voting \citep{choi2025debate}.

\paragraph{Heterogeneous teams and misparsed inputs.}
Rounds 2 and 3 were generated by a harness whose parser misread some answers
written as a mantissa times a power of ten, so in \misTeamPct\% of the
homogeneous and \peerMisPct\% of the heterogeneous teams an agent was shown a
peer value that differed from the one written. Recorded answers are re-scored,
but what agents were shown cannot be corrected. Excluding those teams leaves the
homogeneous results unchanged (total gain \totalGainPctDM\%; genuine peers minus
self-revision $\compPeerSelfDM$ $[\compPeerSelfDMLo, \compPeerSelfDMHi]$) but
removes the significance of the SciOly difference ($\sciPeerSelfDM$
$[\sciPeerSelfDMLo, \sciPeerSelfDMHi]$) and of the degradation of the
heterogeneous pool under genuine peers ($\heteroDebate$ $[\heteroDebateLo,
\heteroDebateHi]$ on all teams; $\heteroDebateDM$ $[\heteroDebateLoDM,
\heteroDebateHiDM]$ after exclusion). We therefore draw no conclusion about
debate in heterogeneous teams.

\section{Interventions (RQ4)}
\label{sec:interventions}

Aggregation and debate leave the item-level component unchanged. We tested
four classes of intervention on the agents and one procedure that uses
information beyond the answers (Table~\ref{tab:interventions}). The primary
outcome is the paired change in team MAE@5; the change in $A$ is reported
alongside.

\begin{table}[t]
\centering\footnotesize
\setlength{\tabcolsep}{3.5pt}
\begin{tabular}{llrrlrr}
\toprule
intervention & against & $N$ & items & $\Delta$ team MAE@5 [95\% CI] & $\Delta A$ (\%) & $p_{\mathrm{Holm}}$ \\
\midrule
\multicolumn{7}{l}{\textit{Composition} (realFP main corpus, 526 items; five models of similar accuracy, $N{=}5$)} \\
mixture of the five & their own teams & 5 & 526 & $\mathbf{-0.142}$ {\scriptsize$[-0.191, -0.095]$} & -- & -- \\
mixture of the five & best member's team & 5 & 526 & $-0.035$ {\scriptsize$[-0.117, +0.046]$} & -- & -- \\
\multicolumn{7}{l}{\textit{Prompt level} (heterogeneous five-model pool, realFP)} \\
per-agent anchors$^{\dagger}$ & baseline & 5 & 529 & $-0.013$ {\scriptsize$[-0.091, +0.067]$} & $-1$ & 0.747 \\
per-agent scaffolds & baseline & 5 & 529 & $\mathbf{-0.104}$ {\scriptsize$[-0.196, -0.015]$} & $-10$ & 0.089 \\
calculator offload & baseline & 3 & 526 & $\mathbf{+0.540}$ {\scriptsize$[+0.363, +0.720]$} & $-52$ & 0.001 \\
offload + anchors$^{\dagger}$ & offload & 3 & 518 & $\mathbf{+0.195}$ {\scriptsize$[+0.017, +0.372]$} & $-14$ & 0.096 \\
offload + reference values & offload & 3 & 507 & $\mathbf{+0.279}$ {\scriptsize$[+0.082, +0.478]$} & $+6$ & 0.037 \\
offload + unit bases & offload & 3 & 516 & $+0.123$ {\scriptsize$[-0.064, +0.312]$} & $-19$ & 0.406 \\
\multicolumn{7}{l}{\textit{Sampling temperature} (four models pooled, realFP, run 0)} \\
$T=0.8$ & $T=0.4$ & 10 & 524 & $-0.013$ {\scriptsize$[-0.030, +0.004]$} & $-3$ & -- \\
$T=1.0$ & $T=0.4$ & 10 & 524 & $-0.005$ {\scriptsize$[-0.024, +0.013]$} & $-4$ & -- \\
\multicolumn{7}{l}{\textit{Activation} (\texttt{qwen2.5-7b}, 150 realFP items, median rule, $\alpha=1$; mean of two jobs)} \\
head masking & none & 5 & 150 & $-0.021$ {\scriptsize$[-0.149, +0.104]$} & $-4^{\ddagger}$ & -- \\
layer skipping & none & 5 & 150 & $\mathbf{-0.276}$ {\scriptsize$[-0.431, -0.126]$} & $-11^{\ddagger}$ & -- \\
residual noise & none & 5 & 150 & $\mathbf{+0.537}$ {\scriptsize$[+0.204, +0.871]$} & $+31^{\ddagger}$ & -- \\
contrastive steering & none & 5 & 150 & $+0.008$ {\scriptsize$[-0.038, +0.053]$} & $+1^{\ddagger}$ & -- \\
\multicolumn{7}{l}{\textit{Information beyond the answers} (realFP main corpus, 15 models; 10-fold cross-validation)} \\
linear recalibration & none & 30 & 529 & $-0.042$ {\scriptsize median; $-0.341$ to $+0.018$} & $+0.5$ & -- \\
\bottomrule
\end{tabular}
\caption{Interventions. $\Delta$ team: paired change in team MAE@5 (OOM; negative
is better) with a 95\% item-clustered interval, bold if it excludes zero.
$\Delta A$: relative change of the item-level component. $^{\dagger}$Anchor
identity is confounded with model identity (Appendix~\ref{app:interventions}).
$^{\ddagger}$Unwinsorized $A$ of the screen, comparable only within the screen.
Winsorization lowers $A$ whenever an arm inflates the within-item spread
severalfold (calculator offload, $B$ $\logcalcDB$~OOM$^2$), so $\Delta A$ is
not interpretable for those arms. Holm: adjusted over the six prompt-level arms.
The recalibration row gives the median and range over models of the change for
teams of 30.}
\label{tab:interventions}
\end{table}

\paragraph{Model composition.}
Across all \nTeams{} mixtures of 2 to 5 of the \nModels{} models, filling five
agent slots and scored on 526 items, a mixture improves on the slot-weighted
mean of its members' homogeneous teams by a median \divGainMed~OOM
(\divGainPct\%; positive for \divGainPos\% of mixtures), about four times the
gain of a homogeneous team of five over one agent (\homoGainFive~OOM). For the
five models of most similar accuracy (single-agent MAE@5 2.08--2.32) the gain is
\divMatched{} [\divMatchedLo, \divMatchedHi]~OOM (\divMatchedPct\%), but the
mixture is not better than the homogeneous team of its best member
($\divMatchedBest$ [$\divMatchedBestLo$, $\divMatchedBestHi$]). Across all
mixtures, \divBeatBest\% improve on their best member by point estimate, and
none improves significantly on the homogeneous team of the most accurate model
(\texttt{\bestHomoModel}, \bestHomoMae); the closest, \bestMixTeam, is worse by
$\bestMixD$ $[\bestMixLo, \bestMixHi]$. Across five-model mixtures, team error
correlates $\predQuality$ with the members' mean accuracy and $\predCorr$ with
their mean pairwise error correlation. Combining architectures therefore
reduces error measurably, as the lower cross-model single-draw correlation
predicts, but in this roster the accuracy differences between models exceed the
gain.

\paragraph{Prompt-level arms.}
In the heterogeneous pool, per-agent scaffolds replace the solve prompt with one
of five decomposition templates per agent. They improve the team by
$\scaffoldDelta$ $[\scaffoldLo, \scaffoldHi]$~OOM (\scaffoldPct\%, $p=\scaffoldP$),
which does not survive Holm correction over the six arms ($p=\scaffoldHolm$) and
is not significant under the median rule; $A$ falls by $\scaffoldDApct$\%
$[\scaffoldDALo, \scaffoldDAHi]$~OOM$^2$ while $B$ rises. Per-agent anchors are
confounded with model identity and are uninformative. Calculator offload, in
which the model lists factors whose product is computed in log space, worsens the
team by $\logcalcDelta$~OOM, and adding reference values worsens it further;
both remain significant after Holm correction: removing arithmetic does not help when the choice of
factors is at fault, and an unconstrained product of several wrong factors is
less bounded than one wrong intuition (catastrophic answers \gridCat\% to
\logcalcCat\%).

\paragraph{Sampling temperature.}
Four models (\texttt{qwen3-4b}, \texttt{qwen3-8b}, \texttt{qwen3-14b},
\texttt{gpt-oss-20b}) were re-run at $T\in\{0.4,0.8,1.0\}$ (run~0, $N=10$).
Raising $T$ to 1.0 increases the dispersion from \BspreadTfour{} to
\BspreadTten{}~OOM. Pooled over the models, team error changes by
$\tPoolEight$ $[\tPoolEightLo, \tPoolEightHi]$~OOM at $T=0.8$ and $\tPoolTen$
$[\tPoolTenLo, \tPoolTenHi]$ at $T=1.0$; the 90\% intervals exclude changes larger
than \tPoolEightMargin{} and \tPoolTenMargin~OOM (\tPoolEightMarginPct\% and
\tPoolTenMarginPct\% of team error). Under the median rule the change at $T=1.0$
is $\tPoolTenM$ $[\tPoolTenMLo, \tPoolTenMHi]$, a gain of about 1\%. $A$ falls by
3--4\% while $B$ rises by a similar amount (Table~\ref{tab:interventions}),
which leaves the error of a team of ten nearly unchanged.

\paragraph{Activation interventions.}
A screen on \texttt{qwen2.5-7b} (150 items, $N=5$, round~1) perturbed
the residual stream at layer 14 of 28, at strengths $\alpha\in\{0.5,1,2\}$ in
units of the layer's root-mean-square activation: contrastive activation
addition along five strategy contrasts (for example top-down against bottom-up
decomposition), applied with opposite signs to pairs of agents; Gaussian noise;
masking of a random subset of attention heads per agent; and skipping layer 14,
the same edit for every agent (Appendix~\ref{app:interventions}). Steering
changed the team answers by more than the variation between jobs of the
unperturbed arm (mean $|\Delta|$ \steerChangeMin--\steerChangeMax~OOM against
\floorChangeMin--\floorChangeMax), but changed neither $\rho$ nor team error.
Skipping layer~14 changed team error by $\skipDeltaMin$ to
$\skipDeltaMax$~OOM in all \skipRuns{} runs (about $-\skipPct$\%) and $A$ by
$\skipDApctMin$ to $\skipDApctMax$\%: in this screen it meets the criterion of
lowering the item-level error and improving the team. It does so by changing the model for every agent, not
by diversifying the agents. The screen stored only per-item team errors computed
with the earlier parser and the median rule, and covers one model and 150 items,
so it identifies a candidate rather than an established effect.

\paragraph{Information beyond the answers.}
A linear map $\log_{10}y\approx a+b\,\bar x_j$ from the item's mean log estimate
to the truth, fitted by least absolute deviations on nine tenths of the items
and applied to the held-out tenth, reduces single-agent error by a median
\recalSoloMed\% (at most \recalSoloMax\%) and the error of teams of thirty by
\recalTeamMed\% of single-agent error (at most \recalTeamMax\%, for
\recalTeamMaxModel), against \recalTeamUncal\% without recalibration. It does so
by removing the overall bias $\mu$: $A$ changes by a median $\recalAMed$\%
(range $\recalAMin$ to $\recalAMax$\%). A recalibration that depends only on the
estimate cannot separate the item-specific error from the true magnitude, so it
does not undo the compression either: the fitted slope $b$ is below one.

\paragraph{Summary.}
Interventions that make the agents differ, through architecture, prompts,
temperature or activation perturbations, increase $B$; only combining
architectures turned this into a consistent gain, and no mixture improved on
the most accurate model. The two arms that improved a team and lowered $A$ were
per-agent scaffolds, which replace the solve template, and layer skipping, which
edits the model; neither is established by the present evidence, the first
because it does not survive correction for multiplicity and the second because
it rests on a screen of one model.

\section{Discussion and limitations}
\label{sec:discussion}

\paragraph{What the estimand says.}
On these benchmarks most of an agent's error is an error that every draw of the
model makes on the item: $\rho$ is \rhoItemMedian{} on realFP, \sciRhoItem{} on
SciOly and \gueRhoItem{} on Guesstimate. Rules that combine the answers, and debate
among agents that answer alike, act on the remainder, and recover a few per
cent of single-agent error. This is a statement about the item distribution of
these benchmarks under these prompts; it does not imply that teams of language
models are equivalent to one agent in general. Where the within-item spread is
large relative to the item-level error, as for one model on Guesstimate and for
the weak models of \citet{chuang2025woc}, aggregation helps, as the decomposition
predicts.

\paragraph{Implications.}
The design question that follows is how to change $c_j$ rather than how to
increase disagreement. In our data the error changed only where the agents changed or
information was added: combining architectures (about 6\% over their own
teams), a different solve template and a model edit (neither established), and
a recalibration that removes the overall bias. Two routes remain untested:
extended test-time reasoning, which our token budget excludes, and
item-specific information, such as retrieval of reference values for each
factor; our reference-value arm supplied a fixed list of constants and was
combined with calculator offload.

\paragraph{Limitations.}
\emph{Scope.} The models are open-weight, 3B--72B, with thinking modes disabled
and \servMaxNew{} new tokens; frontier, closed and reasoning models were not
tested, and the four checkpoints above 15B (two of them quantized) support
consistency with the smaller models rather than a statement about scale.
\emph{Prompts.} The main sweep mixes two prompt pairs unevenly across models;
the second corpus reproduces its structural statistics, but per-model levels of
accuracy and bias, and their rank correlations, are prompt-dependent.
\emph{Revision rounds.} Rounds 2 and 3 were generated with a defective parser
(Section~\ref{sec:debate}); a definitive test of heterogeneous debate requires
regenerating them. \emph{Interventions.} The temperature sweep covers four
models (three of one family), one run and $T\le1$; the activation screen covers
one model, 150 items and the median rule, and its model is recorded in the job
configuration rather than in the stored results; $A$ is sensitive to
winsorization when an arm inflates the spread. \emph{Benchmarks.} SciOly's
answer key offsets its signed errors, and Guesstimate has 15 items.
\emph{Inference.} Only the six prompt-level arms form a confirmatory family;
per-model counts and the remaining contrasts are exploratory.

\paragraph{Reproducibility.}
The prompts are listed in Appendix~\ref{app:prompts} and the models, serving
backends and coverage in Appendix~\ref{app:models}. Every number, table and
figure in this paper is computed from the scored answers by a single analysis
pipeline with fixed random seeds; a complete rerun reproduces them exactly.

\bibliography{refs}
\bibliographystyle{tmlr}

\newpage
\appendix
\counterwithin{table}{section}
\counterwithin{figure}{section}

\section{Prompts}
\label{app:prompts}

Every prompt behind the data, generated from the experimental harness;
$\langle\cdot\rangle$ marks a field filled at run time. A run uses one solve
prompt in round~1 and the matching revise prompt in rounds 2 and~3. SciOly and
Guesstimate use the same prompts as realFP; the temperature sweep, the
heterogeneous pool and the activation screen use the revised pair.

\begin{multicols}{2}

\providecommand{\peerBudget}{120}

\subsection{Prompts of the main and second homogeneous corpora}

\paragraph{Original solve prompt (round~1).}
\begin{list}{}{\setlength{\leftmargin}{1em}\setlength{\rightmargin}{0pt}}
\item[]\scriptsize\ttfamily\raggedright
Give your best order-of-magnitude estimate for this Fermi problem. Reason step by step. Be concise.\\
\mbox{}\\
Problem: $\langle$\textit{question}$\rangle$\\
\mbox{}\\
Format EXACTLY (answer and confidence FIRST, then reasoning):\\
FINAL: \textless{}single number; scientific notation OK, e.g. 3.2e7\textgreater{}\\
CONF: \textless{}0-100\textgreater{}\\
RATIONALE: \textless{}brief reasoning, max 4 lines\textgreater{}
\end{list}

\paragraph{Original revise prompt (rounds 2--3).}
\begin{list}{}{\setlength{\leftmargin}{1em}\setlength{\rightmargin}{0pt}}
\item[]\scriptsize\ttfamily\raggedright
You already solved this problem. Now consider what other team members said.\\
Carefully evaluate their reasoning -- they may be wrong. Update or keep your answer.\\
\mbox{}\\
Problem: $\langle$\textit{question}$\rangle$\\
\mbox{}\\
Your previous answer:\\
$\langle$\textit{own previous answer}$\rangle$\\
\mbox{}\\
Peer responses:\\
$\langle$\textit{peer responses}$\rangle$\\
\mbox{}\\
Format EXACTLY (answer and confidence FIRST, then reasoning):\\
FINAL: \textless{}single number; scientific notation OK, e.g. 3.2e7\textgreater{}\\
CONF: \textless{}0-100\textgreater{}\\
RATIONALE: \textless{}brief reasoning, note any disagreements\textgreater{}
\end{list}

\paragraph{Revised solve prompt (round~1).}
\begin{list}{}{\setlength{\leftmargin}{1em}\setlength{\rightmargin}{0pt}}
\item[]\scriptsize\ttfamily\raggedright
This is a Fermi estimation problem -- the exact value can't be looked up, so estimate it by decomposition.\\
\mbox{}\\
Problem: $\langle$\textit{question}$\rangle$\\
\mbox{}\\
Method:\\
1. Break the quantity into a short chain of factors you can estimate (e.g. population x rate x fraction).\\
2. Give each factor a rough round value WITH its unit.\\
3. Combine them (multiply/divide) into ONE number. Track units and powers of ten carefully -- a\\
~~~dropped or extra factor of 10 is the most common failure.\\
Aim to be within an order of magnitude.\\
\mbox{}\\
Format EXACTLY (answer and confidence FIRST, then reasoning):\\
FINAL: \textless{}your single-number estimate; plain or scientific notation\textgreater{}\\
CONF: \textless{}0-100\textgreater{}\\
RATIONALE: \textless{}the factors and how you combined them\textgreater{}
\end{list}

\paragraph{Revised revise prompt (rounds 2--3).}
\begin{list}{}{\setlength{\leftmargin}{1em}\setlength{\rightmargin}{0pt}}
\item[]\scriptsize\ttfamily\raggedright
You produced an independent estimate for this Fermi problem. Other team members' estimates are below. Your goal is NOT consensus -- the team combines independent estimates, so diverse estimates are valuable; do not drift toward the group just because others differ.\\
\mbox{}\\
Use the peers ONLY to catch a concrete mistake in YOUR OWN work: a wrong unit, a dropped or extra\\
power of ten, an arithmetic slip, or a factor you clearly mis-estimated. If you find such an error,\\
fix it. Otherwise KEEP your estimate.\\
\mbox{}\\
Problem: $\langle$\textit{question}$\rangle$\\
\mbox{}\\
Your previous answer:\\
$\langle$\textit{own previous answer}$\rangle$\\
\mbox{}\\
Peer responses:\\
$\langle$\textit{peer responses}$\rangle$\\
\mbox{}\\
Format EXACTLY (answer and confidence FIRST, then reasoning):\\
FINAL: \textless{}your single-number estimate; plain or scientific notation\textgreater{}\\
CONF: \textless{}0-100\textgreater{}\\
RATIONALE: \textless{}a concrete error you fixed, or why you kept your estimate\textgreater{}
\end{list}

\paragraph{Peer field, genuine peers.}
\begin{list}{}{\setlength{\leftmargin}{1em}\setlength{\rightmargin}{0pt}}
\item[]\scriptsize\ttfamily\raggedright
Agent 1: FINAL: $\langle$\textit{answer}$\rangle$\\
CONF: $\langle$\textit{confidence}$\rangle$\\
TEXT: $\langle$\textit{response, truncated to \peerBudget{} tokens}$\rangle$\\
Agent 2: FINAL: $\langle$\textit{answer}$\rangle$\\
CONF: $\langle$\textit{confidence}$\rangle$\\
TEXT: $\langle$\textit{response, truncated to \peerBudget{} tokens}$\rangle$
\end{list}

\paragraph{Self-revision (revised revise prompt as assembled without peers).}
\begin{list}{}{\setlength{\leftmargin}{1em}\setlength{\rightmargin}{0pt}}
\item[]\scriptsize\ttfamily\raggedright
You produced an independent estimate for this Fermi problem. Other team members' estimates are below. Your goal is NOT consensus -- the team combines independent estimates, so diverse estimates are valuable; do not drift toward the group just because others differ.\\
\mbox{}\\
Use the peers ONLY to catch a concrete mistake in YOUR OWN work: a wrong unit, a dropped or extra\\
power of ten, an arithmetic slip, or a factor you clearly mis-estimated. If you find such an error,\\
fix it. Otherwise KEEP your estimate.\\
\mbox{}\\
Problem: $\langle$\textit{question}$\rangle$\\
\mbox{}\\
Your previous answer:\\
$\langle$\textit{own previous answer}$\rangle$\\
\mbox{}\\
Peer responses:\\
(no peer responses available)\\
\mbox{}\\
Format EXACTLY (answer and confidence FIRST, then reasoning):\\
FINAL: \textless{}your single-number estimate; plain or scientific notation\textgreater{}\\
CONF: \textless{}0-100\textgreater{}\\
RATIONALE: \textless{}a concrete error you fixed, or why you kept your estimate\textgreater{}
\end{list}

\subsection{Prompts of the intervention arms}

\paragraph{Anchors.} Prefixed to the solve and revise prompts; agent slot $i$ receives the $(i \bmod 6)$-th, in this order.
\begin{description}\setlength{\itemsep}{1pt}\scriptsize
\item[\texttt{topdown}] Decomposition strategy: work TOP-DOWN. Start from a large known aggregate (a national or global total, a total population, area, or market size) and scale DOWN by fractions and rates to the target.
\item[\texttt{bottomup}] Decomposition strategy: work BOTTOM-UP. Start from ONE unit (one person, object, or event) and multiply UP by counts and repetitions to reach the target.
\item[\texttt{ratetime}] Decomposition strategy: model the quantity as a RATE x TIME (or flow x duration). Estimate a per-unit-time rate and the span it acts over, then multiply.
\item[\texttt{geometric}] Decomposition strategy: use PHYSICAL DIMENSIONS. Break the quantity into lengths, areas, volumes, or densities (size x size x size, or area x density) and combine.
\item[\texttt{analogy}] Decomposition strategy: ANCHOR TO A REFERENCE. Recall a well-known quantity of the same kind whose size you are confident about, then adjust up or down by explicit ratios to the target.
\item[\texttt{bounds}] Decomposition strategy: BRACKET IT. Estimate a defensible LOWER bound and a defensible UPPER bound independently, then take their geometric midpoint (the square root of their product) as your estimate.
\end{description}

\paragraph{Scaffolds.} Each replaces the whole solve prompt; agent slot $i$ of an item receives the scaffold at position $i$ plus an item-dependent offset in the cycle \texttt{chain}, \texttt{refclass}, \texttt{bracket}, \texttt{dimensional}, \texttt{weinstein}, where \texttt{chain} is the revised solve prompt.

\paragraph{Scaffold \texttt{refclass}.}
\begin{list}{}{\setlength{\leftmargin}{1em}\setlength{\rightmargin}{0pt}}
\item[]\scriptsize\ttfamily\raggedright
Estimate this quantity by COMPARISON, not by building it up from parts.\\
\mbox{}\\
Problem: $\langle$\textit{question}$\rangle$\\
\mbox{}\\
Method:\\
1. Name a quantity of the SAME KIND whose size you are confident about, and state its value.\\
2. Say how the target differs from it -- larger or smaller, and by roughly what ratio.\\
3. Apply that ratio to your reference value.\\
Rest the whole estimate on one anchor and one adjustment. Do not enumerate sub-factors.\\
\mbox{}\\
Format EXACTLY (answer and confidence FIRST, then reasoning):\\
FINAL: \textless{}your single-number estimate; plain or scientific notation\textgreater{}\\
CONF: \textless{}0-100\textgreater{}\\
ANCHOR: \textless{}the reference quantity, its value, and the ratio you applied\textgreater{}
\end{list}

\paragraph{Scaffold \texttt{bracket}.}
\begin{list}{}{\setlength{\leftmargin}{1em}\setlength{\rightmargin}{0pt}}
\item[]\scriptsize\ttfamily\raggedright
BRACKET this quantity rather than guessing it.\\
\mbox{}\\
Problem: $\langle$\textit{question}$\rangle$\\
\mbox{}\\
Method:\\
1. State a value the answer is CERTAINLY ABOVE, and why it cannot be smaller.\\
2. State a value the answer is CERTAINLY BELOW, and why it cannot be larger.\\
3. Tighten whichever bound rests on the weaker argument.\\
4. Report the geometric middle of your final bracket -- the square root of the two multiplied.\\
\mbox{}\\
Format EXACTLY (answer and confidence FIRST, then reasoning):\\
FINAL: \textless{}your single-number estimate; plain or scientific notation\textgreater{}\\
CONF: \textless{}0-100\textgreater{}\\
BRACKET: \textless{}lower bound, upper bound, and why each one holds\textgreater{}
\end{list}

\paragraph{Scaffold \texttt{dimensional}.}
\begin{list}{}{\setlength{\leftmargin}{1em}\setlength{\rightmargin}{0pt}}
\item[]\scriptsize\ttfamily\raggedright
Work BACKWARDS from the units the answer must have.\\
\mbox{}\\
Problem: $\langle$\textit{question}$\rangle$\\
\mbox{}\\
Method:\\
1. State the units the answer must be in.\\
2. Write an equation whose units reduce to those units -- a product or quotient of measurable\\
~~~quantities, SYMBOLS ONLY, no numbers yet.\\
3. Only then assign a magnitude to each symbol.\\
4. Evaluate, and check the units of the result match step 1.\\
\mbox{}\\
Format EXACTLY (answer and confidence FIRST, then reasoning):\\
FINAL: \textless{}your single-number estimate; plain or scientific notation\textgreater{}\\
CONF: \textless{}0-100\textgreater{}\\
EQUATION: \textless{}the symbolic equation, then the magnitudes you substituted\textgreater{}
\end{list}

\paragraph{Scaffold \texttt{weinstein}.}
\begin{list}{}{\setlength{\leftmargin}{1em}\setlength{\rightmargin}{0pt}}
\item[]\scriptsize\ttfamily\raggedright
Estimate this the way a physicist does a back-of-envelope calculation: never put down a quantity you cannot defend.\\
\mbox{}\\
Problem: $\langle$\textit{question}$\rangle$\\
\mbox{}\\
Method:\\
For EVERY quantity you need, do exactly one of these and label it:\\
~~KNOWN   - you are confident of its value; state it.\\
~~BOUNDED - you are not; give a low and a high you can defend, and use their geometric middle.\\
~~DROP    - you cannot even bound it; restructure the estimate so you do not need it.\\
Then combine, and test the result against one independent route.\\
A factor-of-ten answer beats no answer; an undefended digit is worse than an admitted bound.\\
\mbox{}\\
Format EXACTLY (answer and confidence FIRST, then reasoning):\\
FINAL: \textless{}your single-number estimate; plain or scientific notation\textgreater{}\\
CONF: \textless{}0-100\textgreater{}\\
QUANTITIES: \textless{}each quantity tagged KNOWN / BOUNDED / DROP, then the combination and your check\textgreater{}
\end{list}

\paragraph{Calculator-offload solve prompt.}
\begin{list}{}{\setlength{\leftmargin}{1em}\setlength{\rightmargin}{0pt}}
\item[]\scriptsize\ttfamily\raggedright
This is a Fermi estimation problem -- estimate it by decomposition, working in ORDERS OF MAGNITUDE.\\
\mbox{}\\
Problem: $\langle$\textit{question}$\rangle$\\
\mbox{}\\
Method:\\
1. Break the quantity into a chain of factors you can estimate (e.g. population x rate x fraction).\\
2. For EACH factor: give a rough round value WITH its unit and note its power of ten. Check the\\
~~~units cancel to the target unit.\\
3. Do NOT multiply the factors yourself -- a dropped or extra factor of ten is the most common\\
~~~failure. List them on the FACTORS line as numbers joined by * and / (a calculator computes it).\\
Aim to be within an order of magnitude.\\
\mbox{}\\
Format EXACTLY:\\
FACTORS: \textless{}number\textgreater{} * \textless{}number\textgreater{} / \textless{}number\textgreater{} ...   (positive numbers ONLY, joined by * or /; scientific notation OK; no units, no words)\\
FINAL: \textless{}the resulting product as a single number\textgreater{}\\
CONF: \textless{}0-100\textgreater{}\\
RATIONALE: \textless{}each factor with its unit and power of ten, and the unit-cancellation check\textgreater{}
\end{list}

\paragraph{Calculator-offload revise prompt.}
\begin{list}{}{\setlength{\leftmargin}{1em}\setlength{\rightmargin}{0pt}}
\item[]\scriptsize\ttfamily\raggedright
You produced an independent estimate for this Fermi problem. Other team members' estimates are below. Your goal is NOT consensus -- the team combines independent estimates, so do not drift toward the group just because others differ.\\
\mbox{}\\
Use the peers ONLY to catch a concrete mistake in YOUR OWN factors: a wrong unit, a dropped or\\
extra power of ten, or a factor you clearly mis-estimated. Fix it if you find one; otherwise KEEP\\
your estimate. Re-list your (possibly corrected) factors -- do not multiply them yourself.\\
\mbox{}\\
Problem: $\langle$\textit{question}$\rangle$\\
\mbox{}\\
Your previous answer:\\
$\langle$\textit{own previous answer}$\rangle$\\
\mbox{}\\
Peer responses:\\
$\langle$\textit{peer responses}$\rangle$\\
\mbox{}\\
Format EXACTLY:\\
FACTORS: \textless{}number\textgreater{} * \textless{}number\textgreater{} / \textless{}number\textgreater{} ...   (positive numbers ONLY, joined by * or /; scientific notation OK; no units, no words)\\
FINAL: \textless{}the resulting product as a single number\textgreater{}\\
CONF: \textless{}0-100\textgreater{}\\
RATIONALE: \textless{}each factor with its unit and power of ten, and the unit-cancellation check\textgreater{}
\end{list}

\paragraph{Reference constants (prefixed to the solve prompt in the grounding arm).}
\begin{list}{}{\setlength{\leftmargin}{1em}\setlength{\rightmargin}{0pt}}
\item[]\scriptsize\ttfamily\raggedright
REFERENCE CONSTANTS \& BASELINES (Use these exact values when applicable):\\
- World Population: \textasciitilde{}8.1 x 10\textasciicircum{}9 people | US Population: \textasciitilde{}3.4 x 10\textasciicircum{}8 people\\
- Earth Radius: 6,371 km | Earth Surface Area: 5.1 x 10\textasciicircum{}8 km\textasciicircum{}2 | Land Fraction: \textasciitilde{}29\%\\
- Densities: Water = 1,000 kg/m\textasciicircum{}3 (1 g/cm\textasciicircum{}3) | Air = 1.2 kg/m\textasciicircum{}3 | Steel = 7,800 kg/m\textasciicircum{}3\\
- Solar Constant (at Earth orbit): 1,361 W/m\textasciicircum{}2\\
- Time Conversion: 1 Year = 3.15 x 10\textasciicircum{}7 seconds (pi x 10\textasciicircum{}7 s) | 1 Day = 86,400 seconds\\
- Human Averages: Body Mass = 70 kg | Caloric Intake = 2,000 kcal/day | Walking Speed = 1.4 m/s\\
- Unit Conversions: 1 Mile = 1.609 km | 1 Gallon = 3.785 Liters | 1 Pound = 0.453 kg
\end{list}

\paragraph{Unit bases.} Prefixed to the solve prompt; agent slot $i$ receives the entry at position $i$ plus an item-dependent offset, in this order.
\begin{description}\setlength{\itemsep}{1pt}\scriptsize
\item[\texttt{metric}] Unit System Constraint: Reason strictly in SI METRIC units (meters, kilograms, seconds, liters, Joules).
\item[\texttt{cgs\_astro}] Unit System Constraint: Reason strictly in CGS / ASTRONOMICAL units (centimeters, grams, astronomical units, solar masses, ergs).
\item[\texttt{imperial}] Unit System Constraint: Reason strictly in IMPERIAL / US CUSTOMARY units (feet, miles, pounds, gallons, BTUs).
\item[\texttt{analogy\_ref}] Unit System Constraint: Reason strictly by RELATIVE RATIOS to common reference objects (e.g. mass of a Boeing 747, volume of an Olympic pool, height of Eiffel Tower).
\item[\texttt{log\_scale}] Unit System Constraint: Reason strictly in LOGARITHMIC ORDERS OF MAGNITUDE (powers of 10, e.g. 10\textasciicircum{}3 kg, 10\textasciicircum{}6 meters).
\end{description}
\end{multicols}

\FloatBarrier
\section{Models, coverage and prompt versions}
\label{app:models}

\begin{table}[H]
\centering\footnotesize
\begin{tabular}{lllrrr}
\toprule
short name & family & params & precision & ctx & round-1 answers \\
\midrule
\texttt{qwen2.5-3b} & Qwen2.5 & 3.1B & bf16 & 16384 & 182,938 \\
\texttt{smollm3-3b} & SmolLM3 & 3.1B & bf16 & 8192 & 185,677 \\
\texttt{qwen3-4b} & Qwen3 & 4.0B & bf16 & 16384 & 184,977 \\
\texttt{mistral-7b} & Mistral & 7.2B & bf16 & 16384 & 183,330 \\
\texttt{olmo2-7b} & OLMo\,2 & 7.3B & bf16 & 8192 & 185,679 \\
\texttt{qwen2.5-7b} & Qwen2.5 & 7.6B & bf16 & 16384 & 185,683 \\
\texttt{llama3.1-8b} & Llama\,3.1 & 8.0B & bf16 & 8192 & 185,605 \\
\texttt{marin-8b} & Marin & 8.0B & bf16 & 8192 & 185,370 \\
\texttt{qwen3-8b} & Qwen3 & 8.2B & bf16 & 16384 & 184,864 \\
\texttt{phi4-14b} & Phi-4 & 14.7B & bf16 & 16384 & 185,392 \\
\texttt{qwen3-14b} & Qwen3 & 14.8B & bf16 & 8192 & 185,543 \\
\texttt{r1-distill-qwen-14b} & R1-Distill & 14.8B & bf16 & 8192 & 17,295 $^{\dagger}$ \\
\texttt{gpt-oss-20b} & gpt-oss & 21B (3.6B active) & bf16 & 16384 & 184,520 \\
\texttt{qwen3.6-27b} & Qwen3.6 & 27B & bf16 & 8192 & 63,394 \\
\texttt{llama3.3-70b-awq} & Llama\,3.3 & 70.6B & AWQ\,4-bit & 8192 & 96,993 \\
\texttt{qwen2.5-72b-awq} & Qwen2.5 & 72.7B & AWQ\,4-bit & 8192 & 131,774 \\
\bottomrule
\end{tabular}
\caption{Models with Fermi coverage, ordered by parameter count.
$^{\dagger}$Excluded from every statistic. \texttt{gpt-oss-20b} is a mixture of
experts (total and active parameters given); the two \texttt{-awq} models are
4-bit AWQ checkpoints, all others are served in bfloat16 through
\backendName{}, except the activation screen (HuggingFace \texttt{transformers}).
``Round-1 answers'' counts parsed answers.}
\label{tab:models}
\end{table}

\paragraph{Coverage and checks.}
A complete sweep has $117$ agent slots per run ($\sum N$ over the ladder) and
three runs on 529 items; \nCompleteModels{} models are complete, and three
completed fewer runs (\partialModels), so their intervals are wider; item
coverage is complete for all. Round-1 accuracy does not depend on the team size
to which an agent was assigned (median $|e|$ varies by at most
\stableMaxRange~OOM across $N$), except for \texttt{r1-distill-qwen-14b}, which
reaches the token limit on \rOneCapLo--\rOneCapHi\% of generations and is
excluded. Answers are read only from the final answer line; an earlier parser,
which also read the last number in the text and misread mantissa notation, was
replaced and every answer re-scored.

\paragraph{Prompt version.}
The run manifests do not record the prompt pair, but the round-1 prompt length
is fixed by the template for a given model and item and differs by
\promptTokMin--\promptTokMax{} tokens between pairs; every run matches one pair
on all items or on none, except one interrupted run of \promptMixedModels{},
whose answers are assigned individually. Answers up to \promptLastOld{} used the
original pair, from \promptFirstNew{} the revised one. Table~\ref{tab:promptversion}
compares the pairs within the models that ran both.

\begin{table}[H]
\centering\footnotesize
\setlength{\tabcolsep}{3pt}
\resizebox{\textwidth}{!}{\begin{tabular}{lcrrlrrrrrrl}
\toprule
 & runs & \multicolumn{3}{c}{round-1 MAE@5} & \multicolumn{2}{c}{bias} & \multicolumn{2}{c}{IQR, $N{=}5$} & \multicolumn{3}{c}{debate, round 3 $-$ round 1} \\
\cmidrule(lr){3-5}\cmidrule(lr){6-7}\cmidrule(lr){8-9}\cmidrule(lr){10-12}
model & S/D & S & D & D $-$ S & S & D & S & D & S & D & D $-$ S \\
\midrule
\texttt{llama3.1-8b} & 2/1 & 2.99 & 2.63 & $-0.37$ $[-0.46, -0.28]$ & $+1.31$ & $+0.88$ & 0.84 & 0.75 & $+0.063$ & $+0.037$ & $-0.026$ $[-0.071, +0.017]$ \\
\texttt{mistral-7b} & 2/1 & 2.95 & 2.74 & $-0.22$ $[-0.32, -0.12]$ & $+1.38$ & $+1.07$ & 0.03 & 0.14 & $-0.030$ & $+0.025$ & $+0.055$ $[+0.016, +0.097]$ \\
\texttt{phi4-14b} & 2/1 & 2.20 & 2.13 & $-0.06$ $[-0.13, +0.01]$ & $+0.56$ & $+0.58$ & 0.30 & 0.01 & $-0.039$ & $-0.056$ & $-0.021$ $[-0.063, +0.018]$ \\
\texttt{qwen3-14b} & 2/1 & 2.10 & 2.04 & $-0.06$ $[-0.13, +0.01]$ & $+0.40$ & $+0.19$ & 0.00 & 0.00 & $-0.078$ & $+0.018$ & $+0.096$ $[+0.047, +0.150]$ \\
\texttt{qwen3-8b} & 2/1 & 2.16 & 2.08 & $-0.08$ $[-0.16, +0.01]$ & $+0.33$ & $-0.11$ & 0.00 & 0.00 & $-0.050$ & $-0.103$ & $-0.053$ $[-0.091, -0.015]$ \\
\texttt{smollm3-3b} & 1/2 & 2.79 & 2.40 & $-0.39$ $[-0.48, -0.29]$ & $+1.28$ & $+0.69$ & 0.33 & 0.69 & $+0.046$ & $+0.030$ & $-0.016$ $[-0.065, +0.032]$ \\
\midrule
pooled (equal weight) & 11/7 & 2.53 & 2.34 & $-0.20$ $[-0.24, -0.15]$ & $+0.88$ & $+0.55$ & 0.25 & 0.26 & $-0.014$ & $-0.008$ & $+0.006$ $[-0.014, +0.026]$ \\
\bottomrule
\end{tabular}}
\caption{Original (S) against revised (D) prompt pair within models that ran
both, on the same items, with item-clustered intervals for the difference
D$-$S. Bias: mean signed round-1 error, winsorized at $\pm5$ OOM. IQR: the
dispersion, i.e.\ the median within-team interquartile range, at $N{=}5$. Debate: change in team MAE@5 from
round~1 to round~3.}
\label{tab:promptversion}
\end{table}

\FloatBarrier
\section{Supplementary results on single-agent error}
\label{app:round1}

\begin{table}[H]
\centering\footnotesize
\setlength{\tabcolsep}{3.5pt}
\begin{tabular}{lrrrrrrrrrr}
\toprule
 & \multicolumn{3}{c}{coverage} & \multicolumn{3}{c}{error (OOM)} & \multicolumn{4}{c}{accuracy bands (\%)} \\
\cmidrule(lr){2-4}\cmidrule(lr){5-7}\cmidrule(lr){8-11}
model & items & answers & parse & med.\ $e$ & med.\ $|e|$ & MAE@5 & $|e|\!\le\!0.5$ & $|e|\!\le\!1$ & $|e|\!>\!2$ & $|e|\!>\!8$ \\
\midrule
\texttt{qwen2.5-3b} & 529 & 182,938 & 98.5 & $+0.92$ & 2.40 & 2.60\,{\scriptsize$\pm$0.07} & 13.8 & 25.6 & 55.5 & 10.7 \\
\texttt{smollm3-3b} & 529 & 185,677 & 100.0 & $+0.96$ & 2.26 & 2.53\,{\scriptsize$\pm$0.06} & 15.4 & 27.6 & 54.0 & 10.7 \\
\texttt{qwen3-4b} & 527 & 184,977 & 99.6 & $+0.92$ & 2.12 & 2.47\,{\scriptsize$\pm$0.08} & 16.8 & 28.2 & 51.8 & 9.6 \\
\texttt{mistral-7b} & 529 & 183,330 & 98.7 & $+1.50$ & 2.72 & 2.87\,{\scriptsize$\pm$0.07} & 13.6 & 23.2 & 59.8 & 20.1 \\
\texttt{olmo2-7b} & 529 & 185,679 & 100.0 & $+0.30$ & 1.79 & 2.27\,{\scriptsize$\pm$0.07} & 19.3 & 34.2 & 45.7 & 9.9 \\
\texttt{qwen2.5-7b} & 529 & 185,683 & 100.0 & $+0.03$ & 1.92 & 2.32\,{\scriptsize$\pm$0.07} & 19.4 & 33.4 & 48.3 & 9.7 \\
\texttt{llama3.1-8b} & 529 & 185,605 & 100.0 & $+1.46$ & 2.88 & 2.87\,{\scriptsize$\pm$0.07} & 11.9 & 21.9 & 62.8 & 13.2 \\
\texttt{marin-8b} & 529 & 185,370 & 99.8 & $+0.39$ & 2.18 & 2.50\,{\scriptsize$\pm$0.07} & 16.5 & 29.0 & 53.0 & 11.1 \\
\texttt{qwen3-8b} & 527 & 184,864 & 99.5 & $+0.12$ & 1.58 & 2.13\,{\scriptsize$\pm$0.07} & 18.6 & 36.2 & 42.0 & 8.9 \\
\texttt{phi4-14b} & 529 & 185,392 & 99.8 & $+0.42$ & 1.65 & 2.17\,{\scriptsize$\pm$0.07} & 20.5 & 37.1 & 42.8 & 9.8 \\
\texttt{qwen3-14b} & 529 & 185,543 & 99.9 & $+0.30$ & 1.56 & 2.08\,{\scriptsize$\pm$0.07} & 22.9 & 36.7 & 41.8 & 9.2 \\
\texttt{gpt-oss-20b} & 529 & 184,520 & 99.4 & $+0.01$ & 0.95 & 1.69\,{\scriptsize$\pm$0.07} & 34.3 & 52.5 & 30.5 & 8.7 \\
\texttt{qwen3.6-27b} & 527 & 63,394 & 99.2 & $+0.07$ & 1.17 & 1.90\,{\scriptsize$\pm$0.07} & 26.5 & 45.3 & 36.2 & 9.5 \\
\texttt{llama3.3-70b-awq} & 529 & 96,993 & 100.0 & $+0.27$ & 1.39 & 2.02\,{\scriptsize$\pm$0.08} & 24.3 & 40.4 & 40.1 & 8.8 \\
\texttt{qwen2.5-72b-awq} & 529 & 131,774 & 99.7 & $+0.00$ & 1.07 & 1.85\,{\scriptsize$\pm$0.07} & 28.0 & 47.7 & 33.8 & 9.1 \\
\bottomrule
\end{tabular}
\caption{Round-1 accuracy on realFP (main sweep). ``Answers'' is not a sample
size; the item is the unit of inference. MAE@5 with item-clustered standard
error. $|e|>8$: mass outside the binning range.}
\label{tab:r1summary}
\end{table}

\begin{table}[H]
\centering\footnotesize
\begin{tabular}{lrrrr}
\toprule
model & items & slope $\beta_1$ & $r$ & crossing \\
\midrule
\texttt{qwen2.5-3b} & 486 & $-0.429$ & $-0.403$ & $10^{7.2}$ \\
\texttt{smollm3-3b} & 486 & $-0.457$ & $-0.473$ & $10^{7.4}$ \\
\texttt{qwen3-4b} & 484 & $-0.384$ & $-0.372$ & $10^{8.3}$ \\
\texttt{mistral-7b} & 486 & $-0.205$ & $-0.132$ & $10^{16.0}$ \\
\texttt{olmo2-7b} & 486 & $-0.481$ & $-0.519$ & $10^{5.7}$ \\
\texttt{qwen2.5-7b} & 486 & $-0.358$ & $-0.345$ & $10^{5.5}$ \\
\texttt{llama3.1-8b} & 486 & $-0.381$ & $-0.333$ & $10^{9.3}$ \\
\texttt{marin-8b} & 486 & $-0.459$ & $-0.473$ & $10^{6.2}$ \\
\texttt{qwen3-8b} & 484 & $-0.382$ & $-0.427$ & $10^{5.9}$ \\
\texttt{phi4-14b} & 486 & $-0.231$ & $-0.201$ & $10^{8.8}$ \\
\texttt{qwen3-14b} & 486 & $-0.358$ & $-0.386$ & $10^{6.2}$ \\
\texttt{gpt-oss-20b} & 486 & $-0.254$ & $-0.290$ & $10^{6.1}$ \\
\texttt{qwen3.6-27b} & 484 & $-0.216$ & $-0.223$ & $10^{6.3}$ \\
\texttt{llama3.3-70b-awq} & 486 & $-0.335$ & $-0.350$ & $10^{6.7}$ \\
\texttt{qwen2.5-72b-awq} & 486 & $-0.370$ & $-0.398$ & $10^{5.6}$ \\
\bottomrule
\end{tabular}
\caption{Per-model fit of the median signed error on $\log_{10}y$ for
$|\log_{10}y|\le15$; ``crossing'' is the magnitude at which the fitted bias is
zero.}
\label{tab:compression}
\end{table}

\paragraph{Bias by magnitude, lineage and label audit.}
Below $10^6$ all \nModels{} models overestimate and above $10^9$ all
underestimate; the sign is split only in $10^6$--$10^9$, where the models are
already closest to unbiased, so no weighted mixture of them can be unbiased on
small or large quantities. Same-family pairs correlate at $\rSameFam$ against
$\rDiffFam$ for different families (Mann--Whitney $p=\pSameFam$); same-generation
pairs at $\rSameGen$ against $\rDiffGen$ ($p=\pSameGen$). The audit flags an item
when the models agree with each other (interquartile range of their estimates
$\le\auditSpreadMax$~OOM) and disagree with the answer key by at least
\auditOffsetMin~OOM: \auditFlagged{} of \auditItems{} items. Four flagged items
have an independently checkable value, and in each the models are right; most
others are ill-posed. Removing all flagged items changes the pooled median $|e|$
from \robMedAll{} to \robMedKeep{} and the median absolute cross-model mean
error per item from \robAall{} to \robAkeep{}~OOM.

\FloatBarrier
\section{Supplementary results on aggregation}
\label{app:agg}

\begin{table}[H]
\centering\footnotesize
\setlength{\tabcolsep}{3.5pt}
\begin{tabular}{lrrrlrrrr}
\toprule
 & & & & & \multicolumn{2}{c}{given the item} & pooled & \\
\cmidrule(lr){6-7}
model & $N$ & one agent & team & gain [95\% CI] & resampled & unbiased & unbiased & \% realised \\
\midrule
\texttt{gpt-oss-20b} & 30 & 1.693 & 1.602 & $+0.091$ {\scriptsize$[+0.061, +0.123]$} & 1.602 & 0.407 & 0.837 & $7.1$ \\
\texttt{marin-8b} & 30 & 2.499 & 2.374 & $+0.125$ {\scriptsize$[+0.081, +0.171]$} & 2.376 & 0.570 & 0.978 & $6.5$ \\
\texttt{llama3.1-8b} & 30 & 2.881 & 2.793 & $+0.088$ {\scriptsize$[+0.055, +0.124]$} & 2.740 & 0.652 & 1.065 & $4.0$ \\
\texttt{smollm3-3b} & 30 & 2.527 & 2.450 & $+0.077$ {\scriptsize$[+0.049, +0.107]$} & 2.411 & 0.571 & 0.983 & $3.9$ \\
\texttt{qwen2.5-7b} & 30 & 2.323 & 2.266 & $+0.057$ {\scriptsize$[+0.030, +0.087]$} & 2.263 & 0.504 & 0.973 & $3.1$ \\
\texttt{qwen3.6-27b} & 30 & 1.907 & 1.871 & $+0.036$ {\scriptsize$[+0.016, +0.057]$} & 1.865 & 0.434 & 0.904 & $2.5$ \\
\texttt{qwen2.5-72b-awq} & 30 & 1.856 & 1.821 & $+0.034$ {\scriptsize$[+0.008, +0.065]$} & 1.829 & 0.409 & 0.914 & $2.4$ \\
\texttt{qwen2.5-3b} & 30 & 2.610 & 2.566 & $+0.044$ {\scriptsize$[+0.008, +0.081]$} & 2.561 & 0.593 & 1.070 & $2.2$ \\
\texttt{olmo2-7b} & 30 & 2.273 & 2.238 & $+0.035$ {\scriptsize$[+0.012, +0.058]$} & 2.235 & 0.499 & 0.954 & $2.0$ \\
\texttt{phi4-14b} & 30 & 2.176 & 2.150 & $+0.025$ {\scriptsize$[+0.004, +0.046]$} & 2.133 & 0.487 & 1.001 & $1.5$ \\
\texttt{qwen3-4b} & 30 & 2.471 & 2.449 & $+0.022$ {\scriptsize$[+0.002, +0.044]$} & 2.451 & 0.509 & 0.948 & $1.1$ \\
\texttt{mistral-7b} & 30 & 2.870 & 2.858 & $+0.012$ {\scriptsize$[-0.016, +0.041]$} & 2.782 & 0.755 & 1.342 & $0.6$ \\
\texttt{qwen3-8b} & 30 & 2.131 & 2.122 & $+0.009$ {\scriptsize$[+0.000, +0.019]$} & 2.100 & 0.458 & 0.912 & $0.6$ \\
\texttt{qwen3-14b} & 30 & 2.076 & 2.068 & $+0.008$ {\scriptsize$[-0.005, +0.021]$} & 2.053 & 0.450 & 0.881 & $0.5$ \\
\texttt{llama3.3-70b-awq} & 30 & 2.026 & 2.039 & $-0.013$ {\scriptsize$[-0.034, +0.002]$} & 2.033 & 0.433 & 0.902 & $-0.8$ \\
\bottomrule
\end{tabular}
\caption{Teams at the largest $N$ each model ran (geometric mean). ``Gain'': one
agent minus team, with an item-clustered interval. References
(Section~\ref{sec:refs}): MAE@5 of teams whose members are resampled from the
same item's draws, with their signs (``resampled'') or random signs
(``unbiased''), and of teams whose members are drawn across items with random
signs (``pooled''). ``\% realised'': gain as a share of the gain of the unbiased
per-item reference.}
\label{tab:agg}
\end{table}

\begin{table}[H]
\centering\footnotesize
\begin{tabular}{lrrrrrrr}
\toprule
model & $\rho$ & $N_{\mathrm{eff}}(5)$ & $N_{\mathrm{eff}}(30)$ &
$N_{\mathrm{eff}}(\infty)$ & $\sqrt{A}$ (OOM) & $\rho_{\mathrm{item}}$ & run (\%) \\
\midrule
\texttt{qwen2.5-3b} & 0.882 & 1.10 & 1.13 & 1.13 & 2.87 & 0.881 & 0.1 \\
\texttt{smollm3-3b} & 0.867 & 1.12 & 1.15 & 1.15 & 2.76 & 0.789 & 8.1 \\
\texttt{qwen3-4b} & 0.964 & 1.03 & 1.04 & 1.04 & 2.83 & 0.962 & 0.1 \\
\texttt{mistral-7b} & 0.931 & 1.06 & 1.07 & 1.07 & 3.05 & 0.815 & 10.6 \\
\texttt{olmo2-7b} & 0.922 & 1.07 & 1.08 & 1.09 & 2.74 & 0.916 & 0.0 \\
\texttt{qwen2.5-7b} & 0.911 & 1.08 & 1.09 & 1.10 & 2.80 & 0.914 & 0.1 \\
\texttt{llama3.1-8b} & 0.862 & 1.12 & 1.15 & 1.16 & 2.94 & 0.795 & 6.7 \\
\texttt{marin-8b} & 0.814 & 1.18 & 1.22 & 1.23 & 2.73 & 0.805 & 0.1 \\
\texttt{qwen3-8b} & 0.974 & 1.02 & 1.03 & 1.03 & 2.69 & 0.916 & 5.5 \\
\texttt{phi4-14b} & 0.934 & 1.06 & 1.07 & 1.07 & 2.65 & 0.903 & 2.9 \\
\texttt{qwen3-14b} & 0.964 & 1.03 & 1.04 & 1.04 & 2.63 & 0.919 & 4.2 \\
\texttt{gpt-oss-20b} & 0.841 & 1.15 & 1.18 & 1.19 & 2.24 & 0.836 & 0.2 \\
\texttt{qwen3.6-27b} & 0.928 & 1.06 & 1.07 & 1.08 & 2.47 & 0.906 & 0.0 \\
\texttt{llama3.3-70b-awq} & 0.984 & 1.01 & 1.02 & 1.02 & 2.64 & 0.986 & 0.0 \\
\texttt{qwen2.5-72b-awq} & 0.915 & 1.07 & 1.09 & 1.09 & 2.46 & 0.917 & 0.0 \\
\bottomrule
\end{tabular}
\caption{Variance decomposition per model. $\rho$, $N_{\mathrm{eff}}$ and
$\sqrt{A}$ from the fit of $\operatorname{Var}(\bar e_N)=A+B/N$ to the real
teams, which counts the deviation of a run from the model's other runs as
shared; $\rho_{\mathrm{item}}$ from the one-way decomposition of all round-1
draws by item; ``run'': share of the error variance due to the run deviation.}
\label{tab:neff}
\end{table}

\paragraph{The pooled reference.}
Members drawn independently across items with random signs remove both the
item-level error and the difference in difficulty between items; at $N=30$ they
reduce single-agent error by \refRedMin--\refRedMax\%, less than the per-item
reference; the two references answer different questions, and the shortfall of
the real teams is large under both. Under this reference the median share realised is \idealPctMedian\% and the largest
\idealPctMax\%; on SciOly the reference reduction is small for the most accurate
models (\sciRefRedMin--\sciRefRedMax\%), which makes this share unstable, whereas
the per-item reference gives \sciShareItemMed\% (at most \sciShareItemMax\%).

\FloatBarrier
\section{Robustness to the team rule}
\label{app:teamrule}

\begin{table}[H]
\centering\footnotesize
\resizebox{\textwidth}{!}{\begin{tabular}{p{5.6cm}rrrrrrrr}
\toprule
 & \multicolumn{2}{c}{\makecell{realFP\\(main)}} & \multicolumn{2}{c}{\makecell{realFP\\(second)}} & \multicolumn{2}{c}{\makecell{SciOly\\(second)}} & \multicolumn{2}{c}{\makecell{Guesstimate\\(second)}} \\
\cmidrule(lr){2-3}\cmidrule(lr){4-5}\cmidrule(lr){6-7}\cmidrule(lr){8-9}
 & G & M & G & M & G & M & G & M \\
\midrule
error reduction at $N{=}30$, real teams (\% of one agent; median / max) & 1.7 / 5.4 & 1.0 / 5.6 & 1.7 / 6.0 & 1.1 / 5.8 & 1.0 / 5.7 & 1.4 / 10.1 & 2.1 / 21.7 & 1.1 / 35.5 \\
mixtures significantly better than the best homogeneous team & 0/4,928 & 1/4,928 & 0/4,928 & 0/4,928 & 0/4,928 & 0/4,928 & 0/3,458 & 0/3,458 \\
debate: total gain over one agent (\%) & 2.1 & 1.7 & 2.4 & 2.1 & 5.5 & 5.3 & 8.0 & 6.4 \\
debate: models helped / harmed (significant) & 8/4 & 9/3 & 8/2 & 9/3 & 10/0 & 9/0 & 5/3 & 3/2 \\
round 3: genuine peers $-$ self-revision (OOM) & -- & -- & $+0.008$ & $\mathbf{+0.010}$ & $\mathbf{+0.015}$ & $\mathbf{+0.027}$ & $+0.003$ & $+0.012$ \\
heterogeneous pool: genuine peers, round 3 $-$ round 1 (OOM) & $\mathbf{+0.088}$ & $\mathbf{+0.163}$ & -- & -- & -- & -- & -- & -- \\
heterogeneous pool, round 3: genuine $-$ unrelated-item peers (OOM) & $\mathbf{-0.097}$ & $+0.004$ & -- & -- & -- & -- & -- & -- \\
heterogeneous pool: scaffolds $-$ baseline, primary endpoint (OOM) & $\mathbf{-0.104}$ & $-0.092$ & -- & -- & -- & -- & -- & -- \\
heterogeneous pool: scaffolds, Holm-adjusted $p$ (family of 6) & 0.089 & 0.306 & -- & -- & -- & -- & -- & -- \\
temperature contrasts with a significant gain & 0/8 & 2/8 & -- & -- & -- & -- & -- & -- \\
temperature $1.0 - 0.4$, pooled over 4 models (OOM) & $-0.005$ & $\mathbf{-0.026}$ & -- & -- & -- & -- & -- & -- \\
\bottomrule
\end{tabular}}
\caption{Team-level results under the geometric mean (G, the paper's rule) and
the median of the $\log_{10}$ estimates (M). Bold: the 95\% interval excludes
zero.}
\label{tab:teamrule}
\end{table}

The principal conclusions do not depend on the rule: under both, teams of
thirty reduce single-agent error by a few per cent, debate reduces total error
on realFP by \mcGainMainG\% and \mcGainMainM\%, and peers never improve on
self-revision in pooled comparisons. Five results differ in significance. Under
the median, \mcTeamsSigMainM{} of the \nTeams{} mixtures improves significantly
on the best homogeneous team in the main sweep (\mcTeamsSigRepM{} in the second
corpus); the advantage of self-revision over revision with peers on realFP
excludes zero; the pooled temperature gain at $T=1.0$ excludes zero, and two of
the eight per-model temperature contrasts are significant; and the scaffold arm
is not significant. Under the geometric mean, genuine peers are less harmful
than peers from unrelated items in the heterogeneous pool, a difference that
does not survive the exclusion of misparsed inputs. The per-item reference is
weak under the median, because the median of near-identical draws with random
signs does not shrink; we therefore report reductions rather than shares under
that rule.

\FloatBarrier
\section{Supplementary results on debate}
\label{app:debate}

\begin{table}[H]
\centering\footnotesize
\begin{tabular}{lrrrrrrrr}
\toprule
 & \multicolumn{2}{c}{$\sqrt{A}$ (OOM)} & \multicolumn{2}{c}{$W$ (OOM$^2$)} & \multicolumn{2}{c}{$D$ (OOM)} & \multicolumn{2}{c}{team MAE@5} \\
\cmidrule(lr){2-3}\cmidrule(lr){4-5}\cmidrule(lr){6-7}\cmidrule(lr){8-9}
model & r1 & r3 & r1 & r3 & r1 & r3 & r1 & r3 \\
\midrule
\texttt{qwen2.5-3b} & 2.87 & 2.86 & 1.08 & 0.48 & 0.43 & 0.00 & 2.59 & 2.48 \\
\texttt{smollm3-3b} & 2.62 & 2.64 & 1.10 & 0.88 & 0.51 & 0.06 & 2.47 & 2.49 \\
\texttt{qwen3-4b} & 2.84 & 2.82 & 0.31 & 0.22 & 0.00 & 0.00 & 2.46 & 2.41 \\
\texttt{mistral-7b} & 2.85 & 2.85 & 0.83 & 0.74 & 0.06 & 0.00 & 2.84 & 2.82 \\
\texttt{olmo2-7b} & 2.74 & 2.72 & 0.66 & 0.59 & 0.08 & 0.00 & 2.23 & 2.31 \\
\texttt{qwen2.5-7b} & 2.80 & 2.78 & 0.73 & 0.17 & 0.03 & 0.00 & 2.29 & 2.22 \\
\texttt{llama3.1-8b} & 2.83 & 2.82 & 1.45 & 0.68 & 0.78 & 0.10 & 2.80 & 2.82 \\
\texttt{marin-8b} & 2.71 & 2.81 & 1.84 & 1.60 & 0.63 & 0.05 & 2.38 & 2.58 \\
\texttt{qwen3-8b} & 2.61 & 2.56 & 0.20 & 0.26 & 0.00 & 0.00 & 2.13 & 2.07 \\
\texttt{phi4-14b} & 2.61 & 2.59 & 0.53 & 0.21 & 0.18 & 0.00 & 2.16 & 2.10 \\
\texttt{qwen3-14b} & 2.57 & 2.50 & 0.26 & 0.31 & 0.00 & 0.00 & 2.07 & 2.01 \\
\texttt{gpt-oss-20b} & 2.25 & 2.33 & 0.84 & 0.15 & 0.30 & 0.00 & 1.64 & 1.65 \\
\texttt{qwen3.6-27b} & 2.47 & 2.49 & 0.62 & 0.25 & 0.18 & 0.00 & 1.85 & 1.75 \\
\texttt{llama3.3-70b-awq} & 2.64 & 2.61 & 0.11 & 0.14 & 0.00 & 0.00 & 2.04 & 2.01 \\
\texttt{qwen2.5-72b-awq} & 2.45 & 2.45 & 0.50 & 0.46 & 0.00 & 0.00 & 1.82 & 1.81 \\
\bottomrule
\end{tabular}
\caption{Debate at $N{=}5$, main sweep. $\sqrt{A}$: item-level component (one-way
decomposition over all members of all teams); $W$: within-team variance; $D$:
median within-team interquartile range; all on errors winsorized at $\pm5$ OOM
except $D$.}
\label{tab:rounds}
\end{table}

\FloatBarrier
\section{Supplementary results on the interventions}
\label{app:interventions}

\begin{table}[H]
\centering\footnotesize
\begin{tabular}{lrrrrrrrr}
\toprule
arm & $N$ & $A$ & $B$ & $\rho$ & ladder $r$ & team MAE@5 & $|e|>2$ (\%) & parse (\%) \\
\midrule
\texttt{grid} & 5 & 5.12 & 3.98 & 0.563 & 0.791 & 2.26 & 51.5 & 99.7 \\
\texttt{anchor} & 5 & 5.09 & 4.06 & 0.556 & 0.770 & 2.25 & 51.7 & 99.7 \\
\texttt{scaffold} & 5 & 4.62 & 4.53 & 0.505 & 0.753 & 2.16 & 49.4 & 97.9 \\
\texttt{logcalc} & 3 & 2.55 & 10.61 & 0.194 & 0.154 & 2.94 & 66.9 & 93.2 \\
\texttt{anchor\_logcalc} & 5 & 2.47 & 10.68 & 0.188 & 0.178 & 2.89 & 66.9 & 86.7 \\
\texttt{grounding\_logcalc} & 3 & 2.78 & 11.47 & 0.195 & 0.265 & 3.20 & 70.8 & 89.7 \\
\texttt{units\_basis\_logcalc} & 5 & 2.20 & 10.67 & 0.171 & 0.202 & 2.79 & 64.9 & 88.8 \\
\bottomrule
\end{tabular}
\caption{Prompt-level arms of the heterogeneous pool at round~1. $A$, $B$,
$\rho$: one-way decomposition by item (OOM$^2$); ladder $r$: mean correlation
over items of the errors of two agent slots.}
\label{tab:r7arms}
\end{table}

\paragraph{Arms.}
Agent slot $i$ of the pool is served by model $i \bmod 5$. Scaffolds take an
item-dependent offset, so all 25 combinations of model and scaffold occur;
anchors do not, so each model always receives the same anchor and the anchor
arms cannot separate an anchor effect from a model effect. The logcalc arms have
the lowest parse rates (\logcalcParseMin--\logcalcParse\%), because a factor list
that the estimator cannot combine yields no answer.

\begin{table}[H]
\centering\footnotesize
\begin{tabular}{lrrrrrrr}
\toprule
model & $T$ & $A$ & $B$ & $\rho$ & $D$ & one agent & team \\
\midrule
\texttt{qwen3-4b} & 0.4 & 7.09 & 0.432 & 0.943 & 0.000 & 2.162 & 2.128 \\
\texttt{qwen3-4b} & 0.8 & 6.84 & 0.672 & 0.910 & 0.168 & 2.164 & 2.116 \\
\texttt{qwen3-4b} & 1.0 & 6.83 & 0.828 & 0.892 & 0.378 & 2.188 & 2.126 \\
\texttt{qwen3-8b} & 0.4 & 7.08 & 0.268 & 0.964 & 0.000 & 2.085 & 2.083 \\
\texttt{qwen3-8b} & 0.8 & 6.92 & 0.460 & 0.938 & 0.301 & 2.090 & 2.072 \\
\texttt{qwen3-8b} & 1.0 & 6.82 & 0.537 & 0.927 & 0.358 & 2.086 & 2.065 \\
\texttt{qwen3-14b} & 0.4 & 6.74 & 0.429 & 0.940 & 0.000 & 2.046 & 2.021 \\
\texttt{qwen3-14b} & 0.8 & 6.54 & 0.633 & 0.912 & 0.332 & 2.053 & 2.011 \\
\texttt{qwen3-14b} & 1.0 & 6.45 & 0.701 & 0.902 & 0.477 & 2.050 & 2.008 \\
\texttt{gpt-oss-20b} & 0.4 & 4.98 & 1.039 & 0.827 & 0.402 & 1.704 & 1.617 \\
\texttt{gpt-oss-20b} & 0.8 & 4.94 & 1.096 & 0.819 & 0.544 & 1.709 & 1.604 \\
\texttt{gpt-oss-20b} & 1.0 & 4.93 & 1.153 & 0.811 & 0.540 & 1.731 & 1.646 \\
\bottomrule
\end{tabular}
\caption{Temperature sweep, run~0, $N=10$, 529 items. $A$, $B$, $\rho$: one-way
decomposition by item (OOM$^2$); $D$: dispersion (OOM).}
\label{tab:temperature}
\end{table}

\paragraph{Activation screen.}
The screen used \texttt{qwen2.5-7b} at layer 14 of 28 (the middle of the stack),
150 items, $N=5$, round~1, $T=0.4$ and 512 new tokens. For contrastive
activation addition, each direction is the difference between the final-token
hidden states of two framings of one question (top-down against bottom-up,
conservative against generous, reference class against first principles),
normalized to unit length; agents 1 and 2 receive the first direction with
opposite signs, agents 3 and 4 the second, agent 5 the third, and the direction
is added at every position, scaled by $\alpha$ times the layer's root-mean-square
activation. Noise and head masking use a different seed per agent; layer
skipping returns the layer's input for every agent, so $\alpha$ has no effect
and its \skipRuns{} runs are replicates. Two jobs ran at $\alpha=0.5$ and $1$
and one at $\alpha=2$; where two ran, their $\rho$ agree to \repRho{} and their
team errors to \repAgg~OOM. The stored aggregates use the
median team rule and the earlier parser and are not winsorized.

\begin{table}[H]
\centering\footnotesize
\begin{tabular}{lrrrrrl}
\toprule
arm & $\alpha$ & $\Delta A$ (\%) & $\Delta\rho$ & $\Delta$ team & 95\% CI & verdict \\
\midrule
\texttt{headmask} & 0.5 & $-8$ & $-0.082$ & $-0.109$ & $[-0.237, +0.012]$ & lower $\rho$, team n.s. \\
\texttt{headmask} & 1.0 & $-4$ & $-0.007$ & $-0.021$ & $[-0.149, +0.104]$ & no effect \\
\texttt{headmask} & 2.0 & $-2$ & $-0.015$ & $-0.012$ & $[-0.140, +0.112]$ & no effect \\
\texttt{layerskip} & 0.5 & $-11$ & $-0.030$ & $-0.254$ & $[-0.411, -0.104]$ & team better \\
\texttt{layerskip} & 1.0 & $-11$ & $-0.024$ & $-0.276$ & $[-0.431, -0.126]$ & team better \\
\texttt{layerskip} & 2.0 & $-9$ & $-0.010$ & $-0.268$ & $[-0.426, -0.115]$ & team better \\
\texttt{noise} & 0.5 & $-21$ & $-0.202$ & $-0.048$ & $[-0.284, +0.198]$ & lower $\rho$, team n.s. \\
\texttt{noise} & 1.0 & $+31$ & $-0.617$ & $+0.537$ & $[+0.204, +0.871]$ & team worse \\
\texttt{noise} & 2.0 & $-100$ & $-0.967$ & $+1.176$ & $[+0.786, +1.566]$ & team worse \\
\texttt{steer} & 0.5 & $+0$ & $-0.005$ & $-0.007$ & $[-0.060, +0.038]$ & no effect \\
\texttt{steer} & 1.0 & $+1$ & $-0.002$ & $+0.008$ & $[-0.038, +0.053]$ & no effect \\
\texttt{steer} & 2.0 & $-1$ & $-0.002$ & $+0.030$ & $[-0.023, +0.080]$ & no effect \\
\bottomrule
\end{tabular}
\caption{Activation screen, paired against the unperturbed arm of the same job
and strength and averaged over jobs. $\Delta A$: relative change of the
(unwinsorized) item-level component.}
\label{tab:screen}
\end{table}

\FloatBarrier
\section{Per-model results on SciOly and Guesstimate}
\label{app:bench}

\begin{table}[H]
\centering\scriptsize
\setlength{\tabcolsep}{3pt}
\resizebox{\textwidth}{!}{\begin{tabular}{lrrrlrrrrl}
\toprule
model & items & exact OOM (\%) & MAE@5 & mean bias & slope & $\rho$ & $N_{\mathrm{eff}}(30)$ & \% of ref.\ gain & peers $-$ self \\
\midrule
\texttt{qwen2.5-3b} & 499 & 15 & 2.65 & $-0.07$ $[-0.34, +0.18]$ & $-0.44$ & 0.866 & 1.15 & $1.7$ & $+0.044$ $[-0.007, +0.095]$ \\
\texttt{smollm3-3b} & 500 & 18 & 2.49 & $-0.17$ $[-0.41, +0.09]$ & $-0.38$ & 0.890 & 1.12 & $1.3$ & $+0.012$ $[-0.012, +0.038]$ \\
\texttt{qwen3-4b} & 499 & 20 & 2.39 & $-0.24$ $[-0.50, +0.01]$ & $-0.38$ & 0.944 & 1.06 & $0.7$ & $+0.026$ $[-0.009, +0.062]$ \\
\texttt{mistral-7b} & 498 & 19 & 2.60 & $+0.54$ $[+0.27, +0.80]$ & $-0.12$ & 0.890 & 1.12 & $0.5$ & $-0.011$ $[-0.062, +0.039]$ \\
\texttt{olmo2-7b} & 500 & 17 & 2.58 & $-0.23$ $[-0.48, +0.04]$ & $-0.49$ & 0.906 & 1.10 & $1.9$ & $+0.016$ $[-0.050, +0.083]$ \\
\texttt{qwen2.5-7b} & 500 & 20 & 2.28 & $-0.49$ $[-0.72, -0.26]$ & $-0.28$ & 0.906 & 1.10 & $1.1$ & $+0.058$ $[+0.011, +0.105]$ \\
\texttt{llama3.1-8b} & 500 & 19 & 2.44 & $+0.18$ $[-0.07, +0.40]$ & $-0.32$ & 0.839 & 1.18 & $4.8$ & $+0.011$ $[-0.021, +0.045]$ \\
\texttt{marin-8b} & 500 & 17 & 2.57 & $+0.16$ $[-0.09, +0.40]$ & $-0.41$ & 0.815 & 1.22 & $6.5$ & $+0.011$ $[-0.030, +0.050]$ \\
\texttt{qwen3-8b} & 498 & 25 & 2.11 & $-0.50$ $[-0.74, -0.25]$ & $-0.29$ & 0.945 & 1.06 & $1.1$ & $+0.018$ $[-0.006, +0.043]$ \\
\texttt{phi4-14b} & 499 & 25 & 2.02 & $+0.14$ $[-0.08, +0.36]$ & $-0.14$ & 0.884 & 1.13 & $1.3$ & $+0.092$ $[+0.049, +0.137]$ \\
\texttt{qwen3-14b} & 500 & 27 & 2.04 & $-0.45$ $[-0.68, -0.21]$ & $-0.28$ & 0.953 & 1.05 & $0.3$ & $-0.014$ $[-0.035, +0.003]$ \\
\texttt{gpt-oss-20b} & 500 & 46 & 1.13 & $-0.01$ $[-0.14, +0.12]$ & $-0.06$ & 0.704 & 1.40 & $8.1$ & $-0.045$ $[-0.084, -0.010]$ \\
\texttt{qwen3.6-27b} & 498 & 34 & 1.60 & $+0.04$ $[-0.15, +0.21]$ & $-0.00$ & 0.814 & 1.22 & $5.7$ & $+0.005$ $[-0.034, +0.043]$ \\
\texttt{llama3.3-70b-awq} & 500 & 27 & 1.93 & $-0.02$ $[-0.25, +0.21]$ & $-0.22$ & 0.968 & 1.03 & $0.0$ & $+0.006$ $[-0.012, +0.026]$ \\
\texttt{qwen2.5-72b-awq} & 498 & 33 & 1.68 & $-0.27$ $[-0.46, -0.06]$ & $-0.21$ & 0.899 & 1.11 & $3.8$ & $-0.008$ $[-0.028, +0.009]$ \\
\bottomrule
\end{tabular}}
\caption{SciOly, per model (second corpus). Mean bias includes the offset of the
answer key. Slope: of the per-item median signed error on $\log_{10}y$.
\% of ref.\ gain: share of the reduction of the per-item unbiased reference
realised at $N=30$. Peers $-$ self: round-3 team MAE@5 with genuine peers minus
with self-revision.}
\label{tab:benchsci}
\end{table}

\begin{table}[H]
\centering\scriptsize
\setlength{\tabcolsep}{3pt}
\resizebox{\textwidth}{!}{\begin{tabular}{lrrrlrrrrl}
\toprule
model & items & exact OOM (\%) & MAE@5 & mean bias & slope & $\rho$ & $N_{\mathrm{eff}}(30)$ & \% of ref.\ gain & peers $-$ self \\
\midrule
\texttt{qwen2.5-3b} & 15 & 10 & 2.20 & $+1.12$ $[+0.02, +2.05]$ & $+1.91$ & 0.883 & 1.13 & $2.7$ & $+0.547$ $[+0.204, +1.031]$ \\
\texttt{smollm3-3b} & 15 & 26 & 1.52 & $+1.39$ $[+0.86, +1.95]$ & $-1.81$ & 0.633 & 1.55 & $5.9$ & $+0.007$ $[-0.043, +0.060]$ \\
\texttt{qwen3-4b} & 15 & 32 & 0.84 & $+0.63$ $[+0.23, +1.06]$ & $-1.23$ & 0.962 & 1.04 & $2.2$ & $+0.005$ $[-0.014, +0.026]$ \\
\texttt{mistral-7b} & 15 & 10 & 1.54 & $+1.53$ $[+1.19, +1.85]$ & $-0.85$ & 0.707 & 1.40 & $0.3$ & $-0.339$ $[-0.440, -0.249]$ \\
\texttt{olmo2-7b} & 15 & 6 & 1.93 & $+1.92$ $[+1.57, +2.30]$ & $-1.18$ & 0.626 & 1.57 & $0.6$ & $-0.306$ $[-0.464, -0.147]$ \\
\texttt{qwen2.5-7b} & 15 & 25 & 1.10 & $+1.09$ $[+0.78, +1.42]$ & $-0.86$ & 0.715 & 1.38 & $1.5$ & $-0.030$ $[-0.095, +0.039]$ \\
\texttt{llama3.1-8b} & 15 & 9 & 1.96 & $+1.77$ $[+1.18, +2.41]$ & $-0.54$ & 0.548 & 1.78 & $8.8$ & $-0.050$ $[-0.103, +0.004]$ \\
\texttt{marin-8b} & 15 & 3 & 3.31 & $+3.30$ $[+2.88, +3.72]$ & $-0.59$ & 0.476 & 2.03 & $-0.6$ & $+0.113$ $[+0.065, +0.162]$ \\
\texttt{qwen3-8b} & 15 & 42 & 0.76 & $+0.42$ $[+0.00, +0.85]$ & $-1.03$ & 0.949 & 1.05 & $5.4$ & $+0.025$ $[+0.002, +0.049]$ \\
\texttt{phi4-14b} & 15 & 8 & 1.50 & $+1.49$ $[+1.14, +1.86]$ & $-0.76$ & 0.801 & 1.24 & $1.5$ & $+0.070$ $[+0.029, +0.110]$ \\
\texttt{qwen3-14b} & 15 & 11 & 1.45 & $+1.17$ $[+0.60, +1.73]$ & $-0.86$ & 0.907 & 1.10 & $3.0$ & $+0.000$ $[-0.008, +0.007]$ \\
\texttt{gpt-oss-20b} & 15 & 53 & 0.66 & $+0.48$ $[+0.29, +0.64]$ & $+0.19$ & 0.194 & 4.54 & $26.8$ & $-0.070$ $[-0.156, +0.021]$ \\
\texttt{qwen3.6-27b} & 15 & 52 & 0.62 & $+0.14$ $[-0.22, +0.48]$ & $-0.20$ & 0.895 & 1.11 & $1.3$ & $+0.066$ $[-0.007, +0.143]$ \\
\texttt{qwen2.5-72b-awq} & 15 & 67 & 0.41 & $+0.01$ $[-0.22, +0.25]$ & $-0.34$ & 0.887 & 1.12 & $7.9$ & $+0.002$ $[-0.002, +0.005]$ \\
\bottomrule
\end{tabular}}
\caption{Guesstimate, per model; columns as in Table~\ref{tab:benchsci}. With 15
items spanning \gueGtMin--\gueGtMax, intervals are wide and slopes are not
informative.}
\label{tab:benchgue}
\end{table}

\end{document}